\documentclass{emulateapj}

\usepackage{amsmath}
\usepackage{graphicx}
\usepackage{natbib}
\usepackage{upgreek}

\newcommand{\hii}{H~{\sc ii}}

\begin{document}

\title{Collective outflow from a small multiple stellar system}

\author{Thomas Peters\altaffilmark{1}}
\email{tpeters@physik.uzh.ch}

\author{Pamela D. Klaassen\altaffilmark{2}}

\author{Mordecai-Mark Mac Low\altaffilmark{3,4}}

\author{Martin Schr\"{o}n\altaffilmark{4,5}}

\author{Christoph Federrath\altaffilmark{6}}

\author{Michael D. Smith\altaffilmark{7}}

\author{Ralf S. Klessen\altaffilmark{4}}

\altaffiltext{1}{Institut f\"{u}r Theoretische Physik, Universit\"{a}t Z\"{u}rich, Winterthurerstr. 190, CH-8057 Z\"{u}rich, Switzerland}
\altaffiltext{2}{Leiden Observatory, Leiden University, PO Box 9513, 2300 RA Leiden, The Netherlands}
\altaffiltext{3}{Department of Astrophysics, American Museum of Natural History,
79th Street at Central Park West, New York, 
  NY 
10024-5192, USA}
\altaffiltext{4}{Zentrum f\"{u}r Astronomie der Universit\"{a}t Heidelberg,
Institut f\"{u}r Theoretische Astrophysik, Albert-Ueberle-Str. 2,
D-69120 Heidelberg, Germany}
\altaffiltext{5}{Department of Computational Hydrosystems, Helmholtz Centre for Environmental Research - UFZ, Permoserstr. 15, D-04318 Leipzig, Germany}
\altaffiltext{6}{Monash Centre for Astrophysics, School of Mathematical Sciences, Monash University, Vic 3800, Australia}
\altaffiltext{7}{Centre for Astrophysics \& Space Science, University
  of Kent, Canterbury, CT2 7NH, England}

\begin{abstract}
The formation of high-mass stars is usually accompanied by powerful protostellar outflows. Such high-mass outflows are not
simply scaled-up versions of their lower-mass counterparts, since observations suggest
that the collimation degree degrades with stellar mass. Theoretically,
the origins of massive outflows remain open to question because radiative feedback and
fragmentation of the accretion flow around the most massive stars,
with $M > 15$~M$_\odot$, may impede the driving of magnetic disk winds. We here present
a three-dimensional simulation of the early stages of core
fragmentation and massive star formation that includes a subgrid-scale model for protostellar outflows. We find that stars that form in a
common accretion flow tend to have aligned outflow axes, so that the
individual jets of multiple stars can combine to form a collective outflow.
We compare our simulation to observations with synthetic H$_2$ and CO observations and find that
the morphology and kinematics of such a collective outflow resembles
some observed massive outflows, such as Cepheus~A and DR~21.  We finally
compare physical quantities derived from simulated observations of our
models to the actual values in the models to examine the reliability
of standard methods for deriving physical quantities, demonstrating
that those methods indeed recover the actual values to within a factor of 2--3.
\end{abstract}

\maketitle

\section{Introduction}
\label{sec:intro}

Molecular outflows in high-mass star-forming regions appear to differ
from those around low-mass stars not just in their strength, but also
in their lack of collimation. \citet{beuthshep05} suggest that
more massive stars appear to have less collimated outflows, although
conclusive evidence for this awaits high-resolution observations of
massive star-forming regions by ALMA. The nearest massive star-forming
regions with strong outflows, such as DR~21 \citep{davis1996}
and Cepheus~A \citep{narayanan1996} are observed to have
complex jet structures rather than the single, well-collimated jets
typically observed from low-mass stars.

Our published study of the interplay of magnetic fields and self-gravity in the
accretion flow around stars with masses $M_* > 15$~M$_\odot$, which we
call high-mass stars in subsequent discussion, 
suggests that accretion disks around such stars may be vulnerable to
disruption by gravitational torques from their own accretion flows, preventing them
from driving magnetocentrifugal jets at all \citep{petersetal11a}. However, we determined
that observed outflows have substantially higher momentum than either
the large-scale magnetic tower jet driven by the accretion flow
\citep{petersetal11a}, or the ionization-driven outflow \citep{petersetal12b}.

\citet{zinnyork07} review the theoretical controversy over whether
massive stars form by monolithic collapse in isolated cores or
accretion from a cluster environment containing many other stars.
Multiple low- and intermediate-mass stars form in the accretion flow
onto high-mass stars in analytic \citep{krattmatz06} and numerical
models
\citep{bonetal04,smithetal09,petersetal10a,petersetal10c,wangetal10}.
The exceptions to this are numerical models starting with initial core
masses of 100--300~M$_{\odot}$ rather than 1000~M$_{\odot}$
\citep{krumkleinmckee07,comm11,myersetal13}, particularly with strongly centrally
concentrated initial density profiles, which \citet{girietal11} show
to suppress fragmentation regardless of other physical
processes. Indeed, the 
lack of isolated young O stars has long
been noticed observationally \citep{gies1987,gvaramadze2012}, although
whether every OB star formed in a group does remain contested
\citep{oey2012}.

\citet{petersetal12b} suggested that the combined effect of jets driven from the
inner disks of the secondary stars
could explain the observed properties of outflows from massive-star
forming regions, even absent an outflow from the most luminous and
massive star. Jets generally can be approximated to have power proportional to the
masses of their stellar sources \citep{matzner2002}. In a region with
a \citet{salpeter55} initial mass function, the cumulative jet
power will thus be proportional to the cumulative mass, which is
indeed dominated by the low-mass stars regardless of whether the
highest-mass stars actually have a jet or not.  
The collective action of the intermediate-mass stars should begin to dominate
the outflow from the region even before enough mass has accumulated on
the most massive object for it to begin emitting significant amounts
of ionizing radiation, or for all of the low-mass stars to have
formed: \citet{petersetal10a} show that the lower-mass stars form
later than the more massive ones in their simulations.

The combined action of jets was first simulated in a massive collapsing region by
\citet{linaka06} and \citet{nakali07}. This work was extended by \citet{wangetal10}, who coupled
the outflow momentum feedback to sink particles. \citet{cunn11} and \citet{krum12} presented
models with combined radiative and outflow feedback.
These studies included decaying turbulence in the
collapsing gas, but neglected rotation, thus maximizing the support
provided by the outflows against collapse by minimizing their
alignment. However, simulations of massive
  star formation including turbulence show the angular momentum vectors of
  protostars in dense groups are initially closely correlated on length scales of a few tenths of a pc, and only
  become more randomly oriented as the objects grow in mass and
  accrete more distant material \citep{jappkless04,fisher04}. 
Even a model including not only turbulence, but also a better treatment of
  radiative heating, and higher spatial resolution than our model 
 also finds such almost planar structures, extending at least to the
 size of the star forming region in our model of 1500~AU \citep[see
 Fig.\ 2 of][]{krumetal07}. 

We describe a numerical simulation of this early stage of
outflow driving in an initially rotating cloud with turbulence only
induced by its own gravitational instability. Since we do not include
background sources of turbulence, our core collapses
to the center and builds up a rotationally-flattened structure there,
in which the entire stellar group forms, thus minimizing the support, but
maximizing outflow alignment. We expect that more realistic turbulent initial
conditions would lead to fragmentation of our cloud on large scales
and the formation of several collapsing regions with globally
misaligned angular momentum vectors, but at least initially
aligned protostellar spin axes in the densest star-forming regions.
There is observational evidence for outflow alignment from well-separated protostars with
distinct outflows on scales of a few tenths of a pc
in the vicinity of DR21 \citep{davis2007} and in Source~G
of W49A \citep{nsmithetal09}, which has suggested that they formed
from the same flattened rotating cloud.

We compare the result to the observed
outflows from the nearby young, massive star-forming regions Cepheus~A
and DR~21. In later work, we will describe the subsequent development
of the H~{\sc ii} region within the outflow.

In Sec.\ 2 we describe our numerical model, and in Sec.\ 3 we describe
results, including simulated observations.  In Sec.\ 4, we compare our
results to the observations in order to determine the consistency of
our scenario with available evidence.

\section{Numerical Simulation}

We here present the first three-dimensional radiation-hydrodynamical simulation of massive star formation that simultaneously
include feedback by protostellar outflows as well has by heating of both ionizing and non-ionizing radiation. We use
the adaptive-mesh code FLASH \citep{fryxell00} with sink particles \citep{federrathetal10} and our improved version of the
hybrid-characteristics raytracing method \citep{rijk06,petersetal10a}. For more details on our numerical technique
as well as its limitations, we refer to \citet{petersetal10a}.

The initial conditions are identical to our previous simulations \citep{petersetal10a,petersetal10b,petersetal10c,petersetal11a}.
We start from a $1000\,M_\odot$ molecular cloud that has an initial temperature $T = 30\,$K
as well as a core of constant density
$\rho = 1.27 \times 10^{-20}\,$ g\,cm$^{-3}$ within a radius of $r = 0.5\,$pc and then drops
as $r^{-3 / 2}$ out to a radius of $r = 1.6\,$pc. The cloud initially rotates as a solid body with angular velocity $\omega = 1.5 \times 10^{-14}\,$s$^{-1}$.
Magnetic fields are not included in this simulation.
We resolve the collapse of the cloud on the adaptive mesh with a
minimum cell size of $98\,$AU. Sink particles form
at a cut-off density of $\rho_\mathrm{crit} = 7 \times 10^{-16}\,$g\,cm$^{-3}$ and an accretion radius of $r_\mathrm{sink} = 590\,$AU.

The sink particles carry an intrinsic angular momentum (or spin). Their
angular momentum vector changes during accretion of gas such
that the total angular momentum of the system (sink particle and gas)
is conserved. Thus, if $\mathbf{L}_\mathrm{acc}$ is the angular
momentum of the accreted material, $\mathbf{S}$ the intrinsic sink
angular momentum before accretion and $\mathbf{L}$ and $\mathbf{L}'$
the orbital angular momentum of the sink particle before and after
accretion, then the new intrinsic sink angular momentum satisfies
$\mathbf{S}' = \mathbf{S} + \mathbf{L} - \mathbf{L}' +
\mathbf{L}_\mathrm{acc}$.  Note that $\mathbf{L}'$ is
determined by mass, center of mass, and linear momentum conservation during accretion \citep{federrathetal10}.

We launch the protostellar outflows around the sink particles by injecting 10\% of the gas that is currently being accreted
into the cells surrounding the sink particle, as suggested by \citet{koniglpudritz00} as a consequence of the angular momentum
relation in cold, thin disks. The material is added to the cells within a cone of height 1567~AU (or 16 grid cells) and an opening angle 15$^\circ$
aligned to the angular momentum vector of the sink particle. The mass loading of the cells is not homogeneous but smoothed such that it goes to zero
at the boundaries of the cone both as function of radius and angle to avoid sharp density contrasts. We assume a footpoint of 0.5\,AU \citep{rayetal07,pudritzetal07} and
eject the ouflow material with twice the escape velocity $v_\mathrm{esc} = \sqrt{2 G M / r}$ at this radius, which is $v_\mathrm{esc} = 60\,$km\,s$^{-1}$ for
$M = 1\, M_\odot$. The most massive star formed in the simulation reaches a mass of about $M = 10\, M_\odot$. This corresponds to
a maximum outflow velocity of 380\,km\,s$^{-1}$, which is on the
conservative side of the observed range of velocities
\citep[e.g.][]{micono98,coppin98}. 
Based on observational constraints \citep[e.g.][]{bacciotti02,bacciotti04}, we
also transfer 90\% of the angular momentum of the accreted gas to the cells in the outflow cone.
We stress that we do not simply set the velocities within the outflow cone to the outflow velocity, but rather add the outflow momemtum to these cells.
All variability in the outflow is due to variations in the accretion
flow determined from the mass and angular
momentum of accreted gas and to changes in the outflow axis determined
from angular momentum vector of the sink particle.
We impose no additional time dependence on the launching mechanism. 

Our simulation is not isothermal but takes the actual heating-cooling balance between compression and radiative heating on the one hand and
molecular, dust and metal line cooling on the other hand into account. Because of strong shocks, the gas in the outflows can heat up to
$T \gtrsim 10^6\,$K. Such high temperatures lead to a prohibitively small hydrodynamical timestep. To circumvent this problem, we
have artificially enhanced the cooling rates for temperatures in excess of $T \geq 10^{4.3}\,$K. Material below this temperature range, such
as photoionized gas, will stay completely unaffected. The only
dynamical effect of the enhanced cooling rates is to prevent the shock-heated gas
from reaching temperatures well beyond $T \approx 25,000\,$K.

This procedure does remove a lot of thermal energy from the shock. To check whether
this will have an impact on the outflow dynamics, we examine whether the outflowing gas is momentum- or energy-driven.
Let $v_\mathrm{jet}$ be the velocity of the jet material and $n_\mathrm{jet}$ be the H$_2$ number density in the jet.
The cooling time
at the jet termination shock, assuming diatomic gas, is
\begin{equation}
t_c = \frac{7}{5} \frac{k_\mathrm{B} T_s}{n \Lambda(T_s)} ,
\end{equation}
where
\begin{equation}
T_s = \frac{5}{36} \frac{\mu}{k_\mathrm{B}} v_\mathrm{jet}^2
\end{equation}
is the shock temperature for $\gamma=7/5$, $n = 4.6\,n_\mathrm{jet}$ is the number density of free particles assuming fully ionized gas, and 
\begin{equation}
\Lambda(T_s) \sim 10^{-22} \frac{\text{erg\,cm}^3}{s} \left(\frac{T_s}{10^6\,\mathrm{K}}\right)^{-0.7} .
\end{equation}
The last relation is valid over the range of temperatures of roughly
$10^{4.5}$~K to $10^{6.5}$~K for solar metallicity
\citep{maclowmccray88}. Here, $k_\mathrm{B}$ is Boltzmann's constant,
$\mu = 2.14 m_\mathrm{p}$ is the mean molecular weight with the proton
mass $m_\mathrm{p}$. Plugging in some typical numbers, $v_\mathrm{jet} =
100\,$km s$^{-1}$ and $n_\mathrm{jet} = 10^{-20}\,$cm$^{-3}$, we
arrive at 
$t_c \approx 3 \times 10^7\,$s. On the other hand, the
crossing time of the jet is $t_x = L_\mathrm{jet} / v_\mathrm{jet}$
with the jet length $L_\mathrm{jet}$. For a snapshot near the
beginning of the simulation, $L_\mathrm{jet} = 0.4\,$pc, we have $t_x
\approx 10^{11}\,$s and thus $t_x \gg t_c$. The difference between
these two numbers is so large that $t_x \gg t_c$ still holds when
slightly different representative numbers and later snapshots are
considered. Hence, we can safely conclude that the outflow is
physically completely in the
momentum-driven regime so that the enhanced cooling does not further impact
its dynamics.

We here report results from the first part of the simulation, before the stars reach masses larger than 10\,$M_\odot$. At these early times,
the \hii\ regions only extend to a few grid cells around the sink particles. The adaptive mesh at the end of this part of the simulation
has 487 million grid cells, and the simulation has consumed 1.73 million CPU hours.

\section{Analysis}

We start our analysis with a discussion of the stellar group that
forms during the simulation and the associated outflow feedback
(Section~\ref{sec:accrhist}). 
We then analyze the collective outflow that forms from the
interactions of the individual
stellar outflows in Section~\ref{sec:outevo}. In
Section~\ref{sec:synobs}, we present 
synthetic observations of this outflow.

\subsection{Accretion History and Outflow Feedback}
\label{sec:accrhist}

\begin{figure*}
\centerline{\includegraphics[height=170pt]{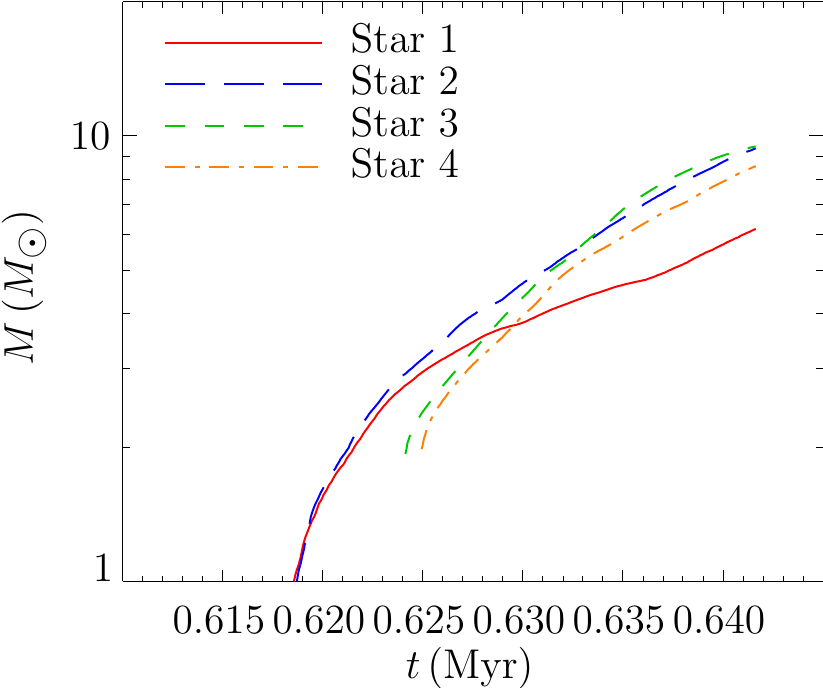}
\includegraphics[height=170pt]{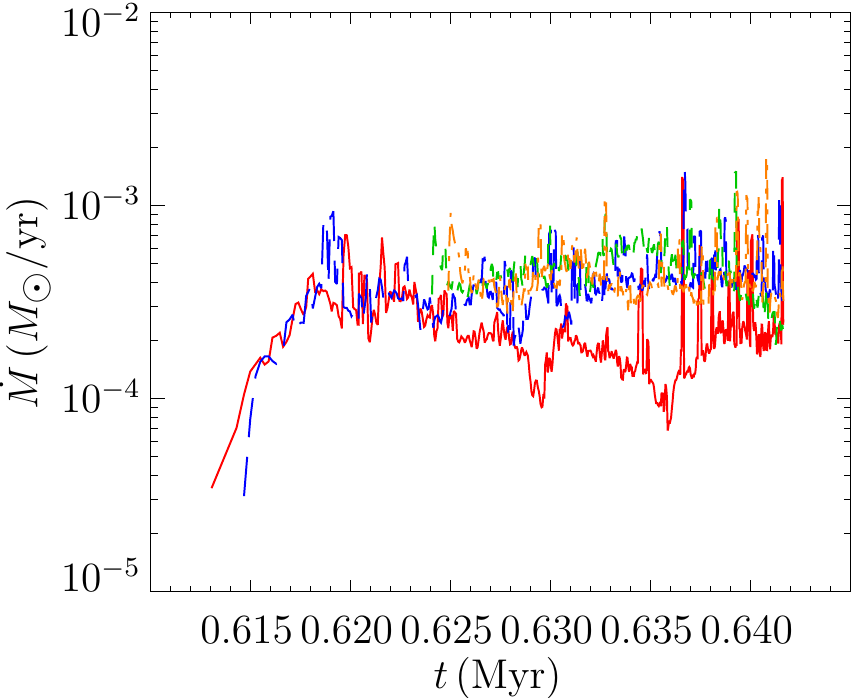}}
\caption{Accretion history of the stellar group. The figure shows the mass ({\em left}) and accretion rate ({\em right})
of the four stars that form during the simulation runtime. Three of the four stars accrete at a relatively constant rate of
$\sim 4\times 10^{-4} M_\odot$\,yr$^{-1}$, whereas the first star that forms accretes at a somewhat lower rate of $\sim
2\times 10^{-4} M_\odot$\,yr$^{-1}$.}
\label{fig:accrhist}
\end{figure*}

The accretion history of the stellar group is shown in Figure~\ref{fig:accrhist}. Four stars form during the simulation runtime, three of
which are close to 10\,$M_\odot$. These three stars accrete at an average rate of $\sim 4\times 10^{-4} M_\odot$\,yr$^{-1}$, while
the first star that forms accretes at the slightly lower rate of $\sim 2\times 10^{-4} M_\odot$\,yr$^{-1}$.

\begin{figure*}
\centerline{\includegraphics[height=170pt]{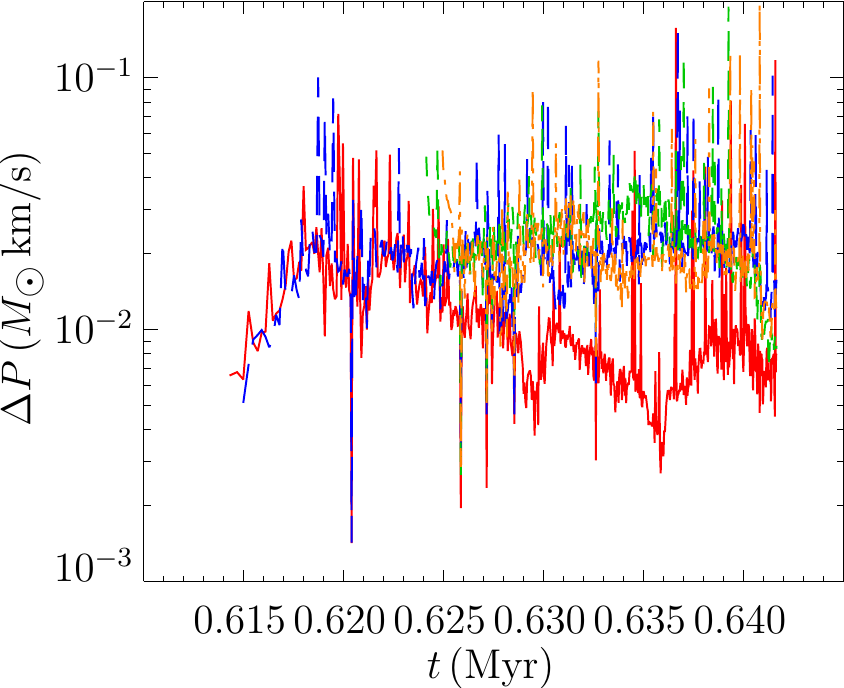}
\includegraphics[height=170pt]{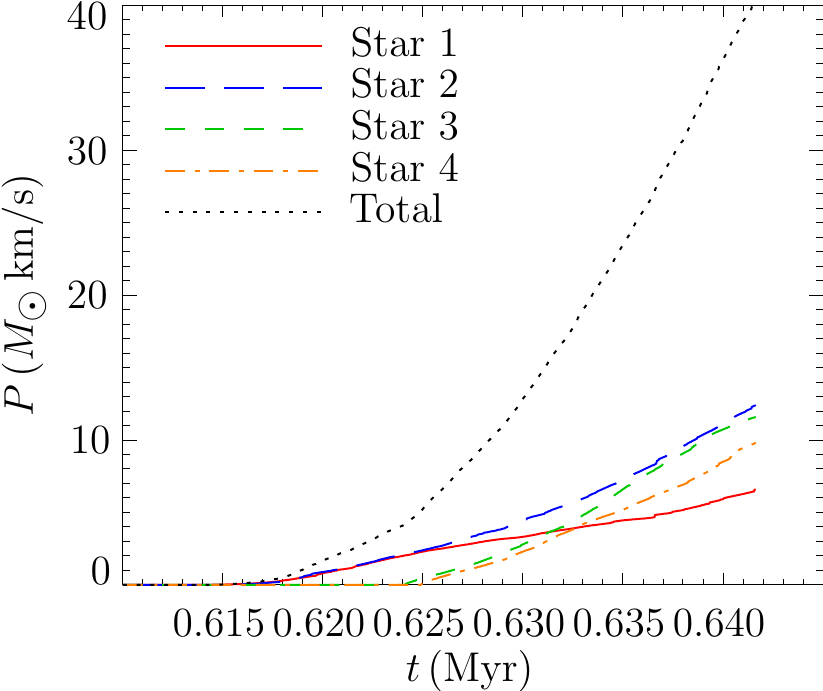}}
\caption{Momentum feedback by the protostellar outflows. The figure shows the instantaneous ({\em left}) and cumulative ({\em right})
momentum imparted on the gas near the stars by the outflows. The cumulative plot also shows the total momentum of all four outflows.}
\label{fig:momentum}
\end{figure*}

The momentum injected into the cloud by the protostellar outflows is shown in Figure~\ref{fig:momentum}. We plot both the
instantaneous momentum $\Delta P$ at a single time step and the cumulative momentum $P$ integrated over the age of the star.
All four stars make a similar contribution
to the total momentum of the collective outflow. The lower accretion rate and consequently lower mass of the first star results in a
smaller contribution compared to the other three stars, so that the cumulative momentum output of the first star is only half of the momentum
output of each of the other three stars. The fluctuations of the accretion rate result in a related variability of the instantaneous
outflow momentum. This effect dominates the magnitude of the instantaneous outflow feedback, so that the outflow from a more
massive star is not necessarily stronger than for a lower-mass star, although the escape velocity scales with stellar mass.

In our setup, all stars form in a rotationally flattened structure or toroid in the center of the simulation box \citep{petersetal10a,petersetal10c}.
The stars revolve around the center of rotation while driving the outflows. Since the outflow cones of the different stars furthermore
partially overlap, the momentum transport within the outflow is quite complex. The fact that one of the stars is injecting a little
less momentum into the outflow than the other stars does not mean that a region in the outflow that moves slower than
the rest can be identified.
Furthermore, the momentum in the outflow is redistributed through shocks that form when the individual outflows collide (see Section~\ref{sec:outevo}).

Directly after their formation, the initial angular momenta $\mathbf{S}_0$ of the sink particles
point in the same direction as the disk
angular momentum vector, which is oriented
perpendicular to the disk. Later, as more
gas falls onto the disk, gravitational instability sets in and leads to the formation of dense filaments. The stars are embedded
in these filaments, which extend significantly above and below the disk, so the accreted gas has a preferred location with respect
to the sink particles that is different from the disk midplane. Since the filaments are generally denser than the rest of the disk, the material
of the filaments dominates the angular momentum balance of gas accreted onto the sink particles. Continuous accretion of gas above or below
the disk plane results in a systematic trend for
the angular momentum $\mathbf{S}$ of the sink particle. Over time, the favored accretion from a certain direction changes the direction of the sink angular
momentum vectors significantly.

\begin{figure*}
\centerline{\includegraphics[height=170pt]{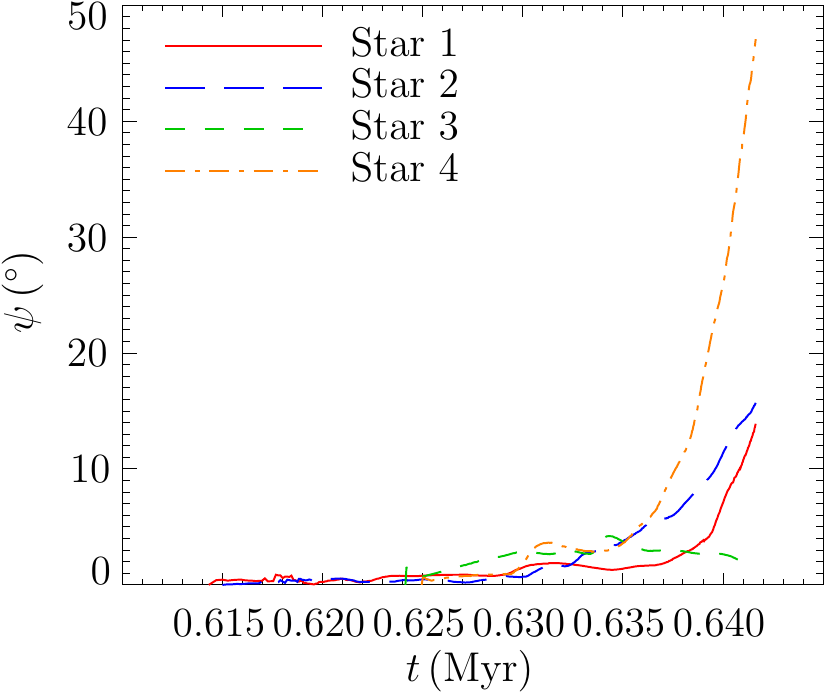}}
\caption{Angle $\psi$ between the initial and the current angular momentum vector of the sink particles as function of time. As the disk
becomes gravitationally unstable and vertically extended filaments form, the angular momentum vectors become misaligned with the
disk angular momentum.}
\label{fig:anglepsi}
\end{figure*}

We quantify the changes in the sink angular momenta by looking at the angle $\psi$ between the initial and the current angular momentum
vector, $\psi = \arccos(\mathbf{j}_0 \cdot \mathbf{j})$ with $\mathbf{j}_0 = \mathbf{S}_0 / \left|\mathbf{S}_0\right|$ and
$\mathbf{j} = \mathbf{S} / \left|\mathbf{S}\right|$. Since $\mathbf{j}_0$ is orthogonal to the disk plane, $\psi$ can be interpreted as an
inclination angle. Figure~\ref{fig:anglepsi} shows $\psi$ as a function of time for all sink particles. Three of the four sink angular momentum
vectors begin to wander after ($t = 0.635\,$Myr), and one of them
reaches a $\psi$ of almost 50$^\circ$ by the end of the simulation.

\begin{figure*}[t]
\begin{center}
\includegraphics[width=400pt]{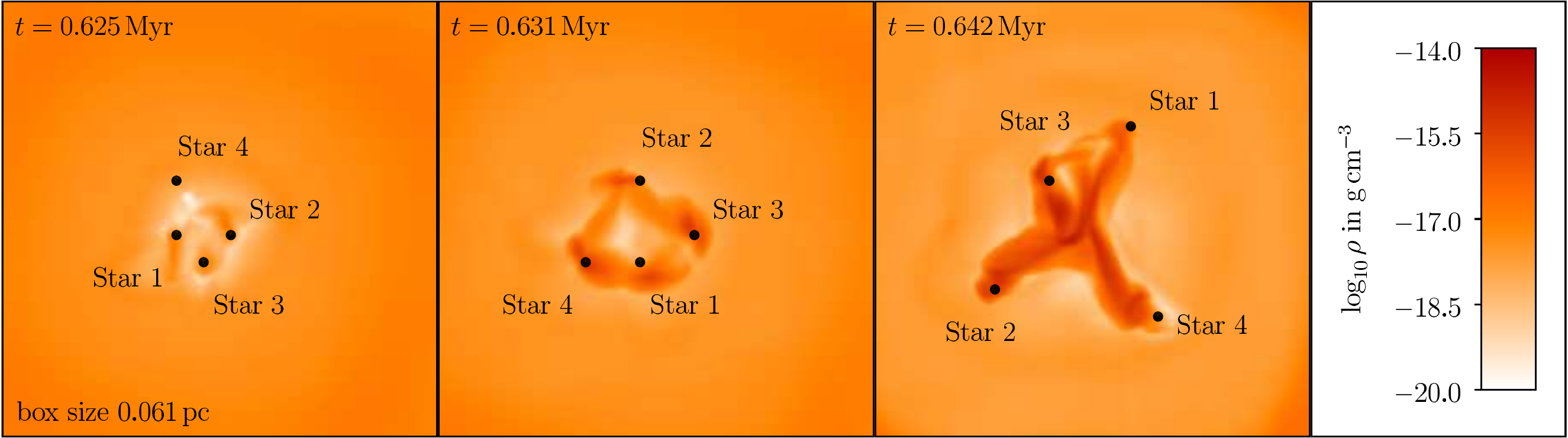}
\end{center}
\vspace{-10pt}
\caption{Density slices through the disk plane for three different snapshots. As more and more gas falls onto the disk, gravitational instability sets in and filaments form.}
\label{fig:faceon}
\end{figure*}

\begin{figure*}[t]
\begin{center}
\includegraphics[width=400pt]{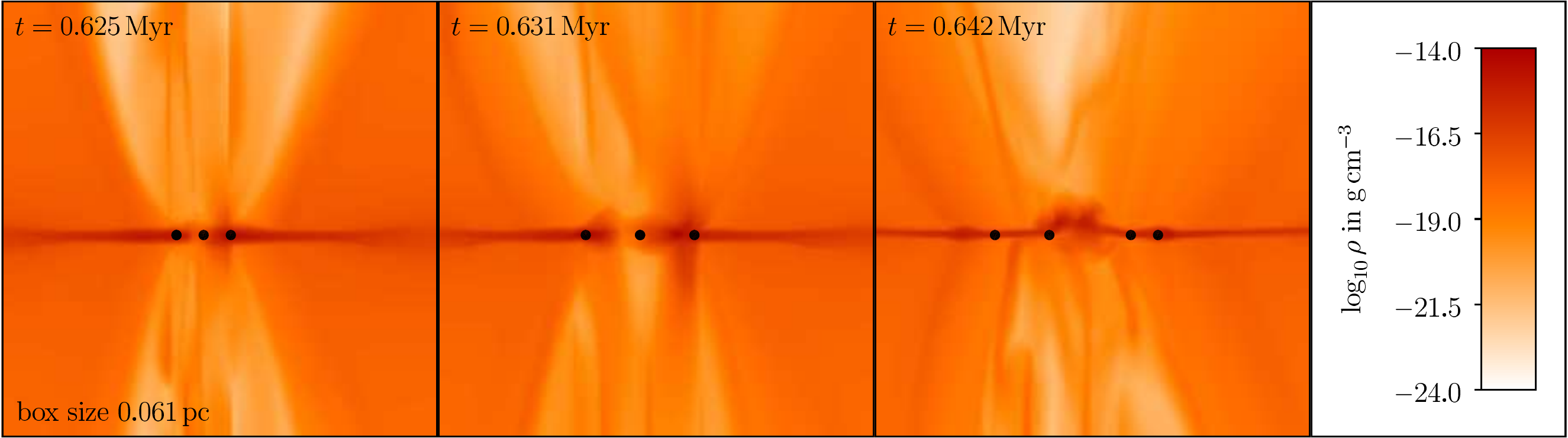}
\end{center}
\vspace{-10pt}
\caption{Density slices perpendicular to the disk plane for the same snapshots as in Figure~\ref{fig:faceon}. One can discern the accumulation
of gas above and below the disk as time proceeds.}
\label{fig:edgeon}
\end{figure*}

The formation of the filaments in the disk is displayed in Figures~\ref{fig:faceon} and~\ref{fig:edgeon}. The figures shows snapshots
from the time when all sink particle
angular momenta are still aligned perpendicular to the disk ($t = 0.625\,$Myr), the time when the first angular momentum vectors begin
moving ($t = 0.631\,$Myr) and the very end of the simulation ($t = 0.642\,$Myr) in face-on and edge-on slices, respectively.
The correlation between the formation of filaments
and the departure of the inclination angle $\psi$ from $\psi = 0^\circ$ is clearly visible. 

\subsection{Evolution of the Collective Outflow}
\label{sec:outevo}

\begin{figure*}[t]
\begin{center}
\includegraphics[width=400pt]{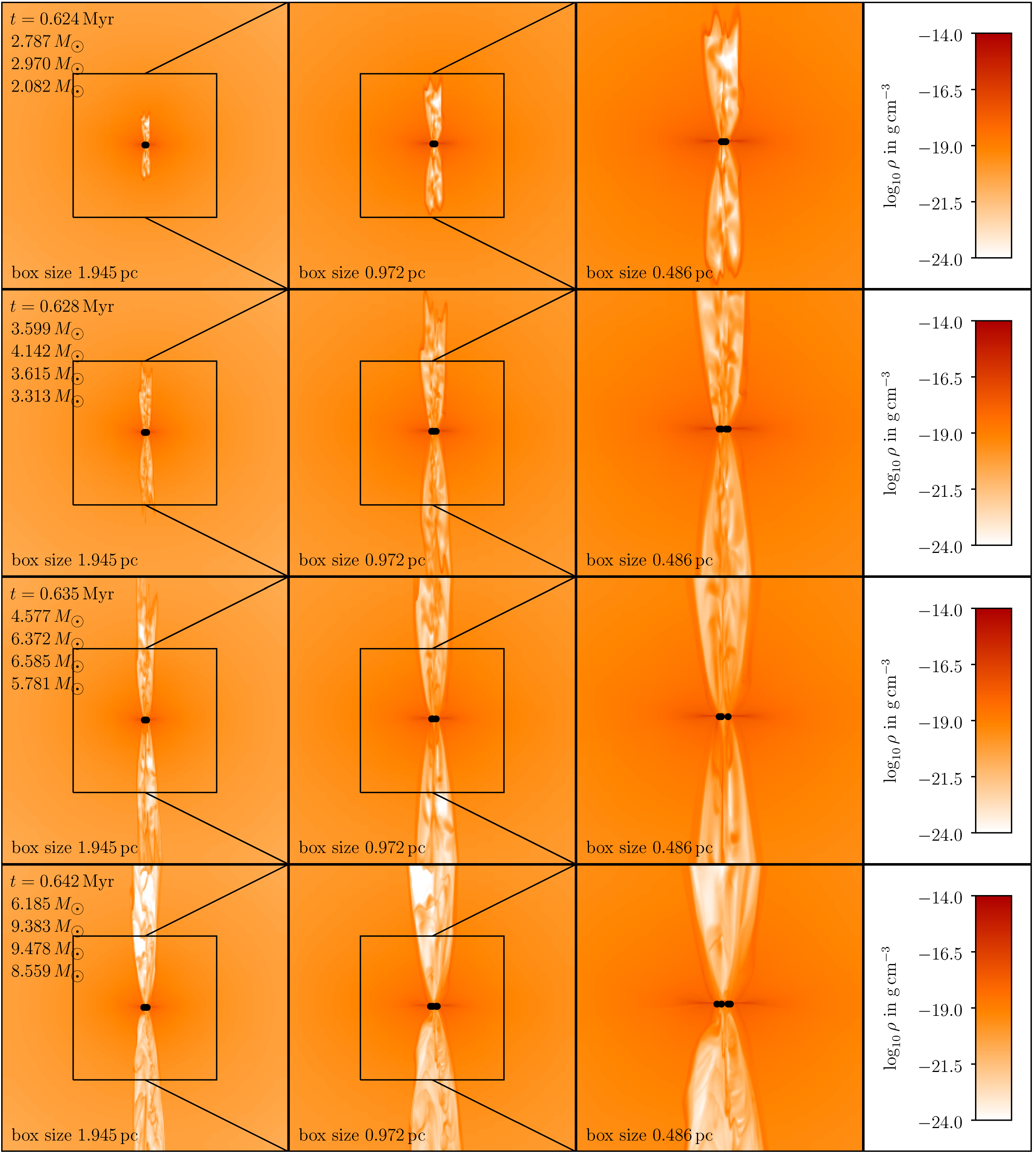}
\end{center}
\vspace{-20pt}
\caption{Density slices through the collective outflow at four different snapshots (rows) and with three different magnification factors (columns).
The masses of the stars in the group are given in the image.
An animated version of this figure is available in the online material for this article.}
\label{fig:panelplot}
\end{figure*}

We show the time evolution of the collective outflow in Figure~\ref{fig:panelplot}. The outflow develops immediately after the first
star forms at $t = 0.613\,$Myr and its leading edge leaves the simulation box at $t = 0.635\,$Myr. In the middle of this time interval,
at $t = 0.624\,$Myr, the outflow only extends to a quarter of the simulation box, which shows that the outflow growth is not
linear in time. The collective outflow is driven by the momentum of all the individual
outflows, and this outflow feedback is highly episodic due to the
coupling of the outflow momentum to the current accretion rate, whose
strong variation is shown in Figure~\ref{fig:momentum}. We therefore do not expect a simple relation between the size of the outflow and the time
since its formation. 

As can be seen in Figure~\ref{fig:panelplot}, the quickly moving shock front (strong density enhancement) at the tip of the outflow is subject to instabilities
that lead to the formation of finger-like structures. The gas density inside the outflow shows substantial variations
between $10^{-24}$ and $10^{-19}\,$g\,cm$^{-3}$. The sudden increase of the accretion rate onto one of the stars leads to intensified
outflow feedback, which in turn triggers the formation of a shock front, which is visible as a density enhancement in the interior
of the outflow. This shock then propagates through the entire outflow until it dissipates
at the outflow boundary or leaves the simulation box. The movie provided in the online material shows that the variation of the outflow
angle and the movement of the stars in the disk leads to a significant variation in the way the outflow material is injected near the disk plane.
In particular, the collision of shocks launched from different stars and moving at different speeds or encountering each other at acute angles
can be observed frequently during the entire evolution.

The collimation of the outflow decreases slightly with time. This observation can be
related to the increasing growth of the inclination angle of the individual outflows (compare Figure~\ref{fig:anglepsi}).
As the individual outflow inclinations rise, the effective opening angle of the collective outflow becomes larger,
extending the base of the collective outflow.

\begin{figure*}[t]
\begin{center}
\includegraphics[width=400pt]{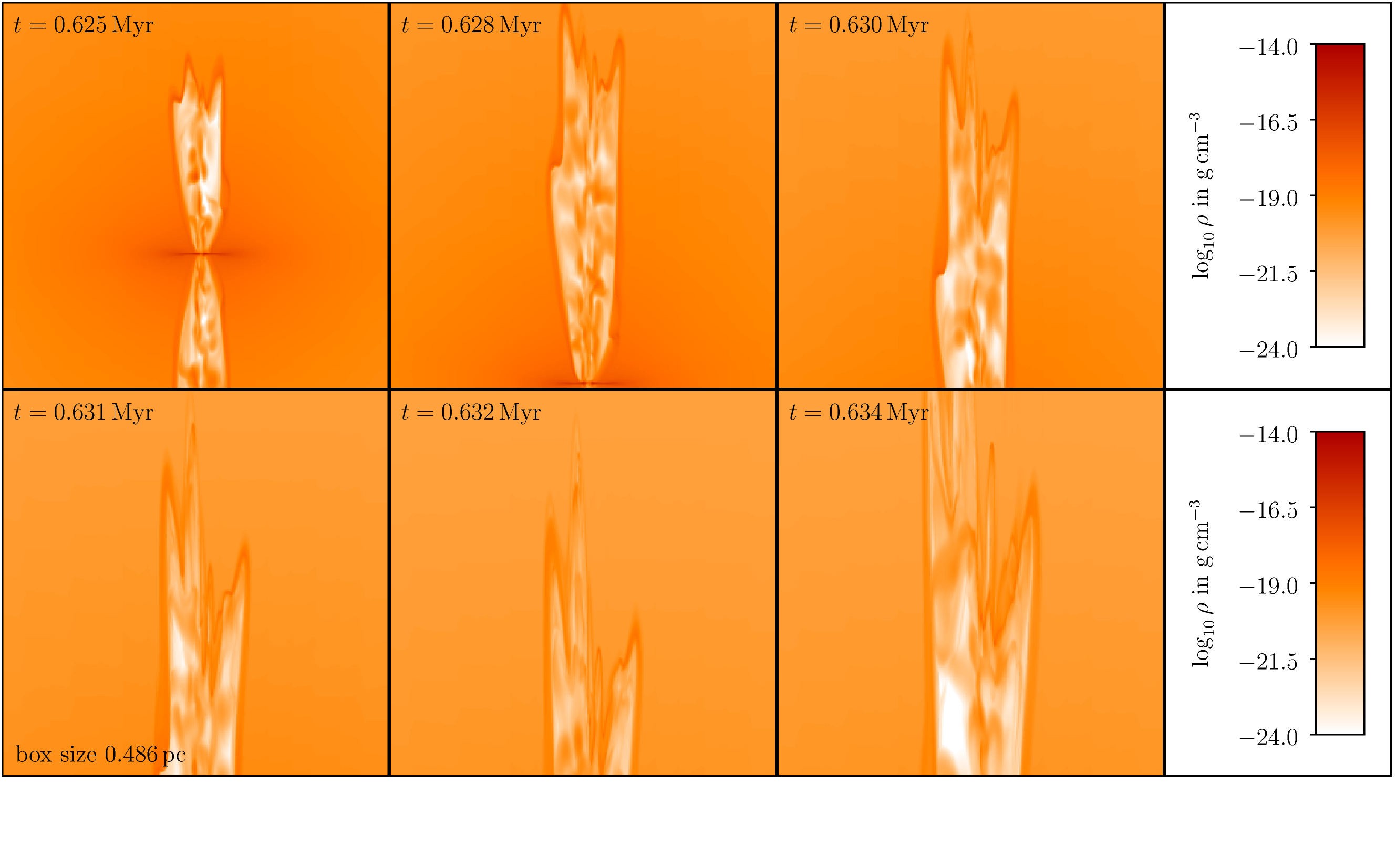}
\end{center}
\vspace{-20pt}
\caption{Density slices through the outflow head at six different snapshots. The different launching directions of the individual
outflows lead to spatially separated outflow tips that form the head of the collective outflow.}
\vspace{25pt}
\label{fig:head}
\end{figure*}

The head of the collective outflow has a distinctive
morphology. Figure~\ref{fig:head} shows the time evolution of one of
the outflow heads until the outflow head leaves the simulation box.
The movie in the online material shows that the multiple heads
continuously come and go during the evolution of the jet.  Their
origin becomes clear once one realizes that the velocity of the jet
increases as it propagates down the density gradient of the envelope,
by roughly a factor of three from early times to when the head leaves
the box. The accelerating interface between the lower density jet gas
and higher density envelope gas has an effective gravity with a sign
opposite to that of the acceleration, so that the interface is Rayleigh-Taylor
unstable.  The head structure thus reflects the usual Rayleigh-Taylor
bubble and spike morphology \citep[e.g.][]{sharp1984}.
As the minimum wavelength of this instability is determined only by
viscosity, higher-resolution simulations will show finer small-scale
structure until the effects of magnetic fields and physical viscosity
are included and resolved. \citet{fujitaetal09} show an example
of this phenomenon in a blowout-induced Rayleigh-Taylor instability.

\subsection{Synthetic Observations}
\label{sec:synobs}

We have generated synthetic observations of our simulated outflows to compare our results with previous simulations and observations. We present
H$_2$ maps of the outflow in Section~\ref{sec:synobsh2} and CO observations in Section~\ref{sec:synobsco}. We have used our customized version
of the three-dimensional adaptive-mesh radiative transfer code RADMC-3D\footnote{http://www.ita.uni-heidelberg.de/$\sim$dullemond/software/radmc-3d/}
to simulate the observations.

\subsubsection{H$_2$ Emission Maps and Spectra}
\label{sec:synobsh2}

We study emission of the collective outflow in the H$_2$ S(1) 1-0 line at $2.1218\,\upmu$m. To this end, we employ the method developed
by \citet{suttner97}. This scheme assumes statistical equilibrium, but not local thermodynamic equilibrium, for the vibrational
states, and solves the full set of rate equations for the three level system consisting of the lowest energy states.
     To determine the H$_2$ density, we 
assume statistical equilibrium
between H$_2$ formation on dust grains and the two temperature-dependent collisional dissociation processes
$\mathrm{H}_2 + \mathrm{H} \to \mathrm{H} + \mathrm{H} + \mathrm{H}$ and
$\mathrm{H}_2 + \mathrm{H}_2 \to \mathrm{H}_2 + \mathrm{H} + \mathrm{H}$.
We use the dust formation rate from \cite{holmck79}, and for the destruction processes
we employ the interpolation formula from \cite{glovmcl07}, which
interpolates between the low-density (\citealt{lepshl87,shpkng87}) and high-density (\citealt{mclshl86,mkm98})
limits of these rates.
However, at shock fronts, although we account for collisional
dissociation, we do not account for shock photodissociation, so our
emission levels are likely upper limits at fast shocks such is in our
jet heads.

\begin{figure*}[t]
\begin{center}
\includegraphics[width=400pt]{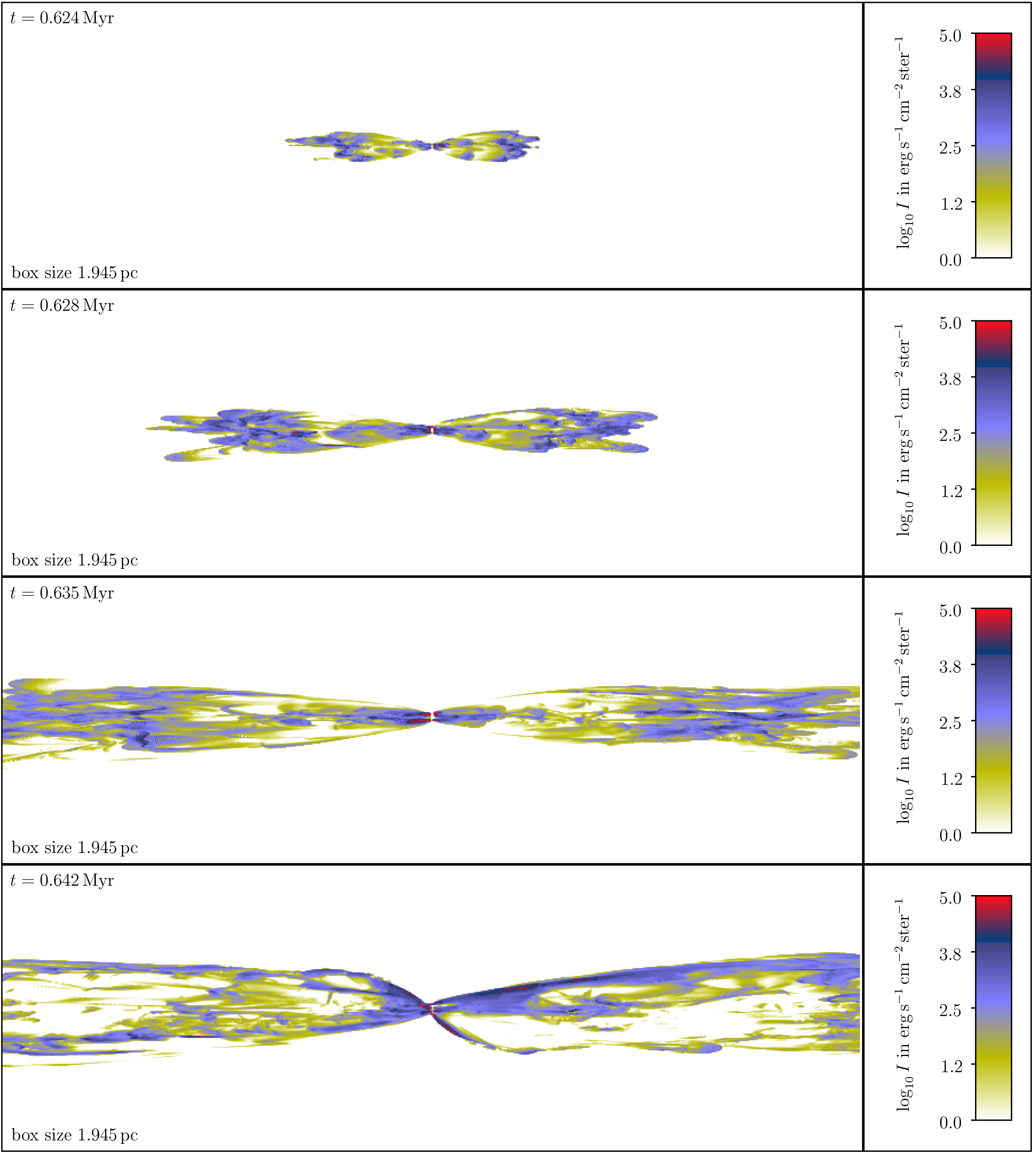}
\end{center}
\vspace{-20pt}
\caption{Intensity of the H$_2$ S(1) 1-0 line for the four snapshots of Figure~\ref{fig:panelplot}. The spatial scale is the same in all four images.
An animated version of this figure is available in the online material for this article.}
\label{fig:panelplotH2}
\end{figure*}

Figure~\ref{fig:panelplotH2} shows the H$_2$ emission for the four snapshots displayed in Figure~\ref{fig:panelplot}. The spatial
scale of the four images is identical and corresponds to the left panels of Figure~\ref{fig:panelplot}.
It can be seen that the extent of the outflow is well represented by the density slices of Figure~\ref{fig:panelplot},
but that multiple tips are present at the outflow head that are not visible in the selected slice plane. This is not surprising because
the slice plane lies perpendicular to the disk that harbors the four stars (compare Figure~\ref{fig:faceon}). Since the H$_2$ emission is optically thin, the H$_2$ maps
reveal the three-dimensional structure of the outflow better than the density slices. However, the emission from different regions
can overlap, leading to confusion along the line of sight.

We can compare these maps to the three-dimensional outflow simulations by \citet{voelker99}. Their Hammer jet model includes precession
of the outflow launching angle, a sinusoidal perturbation of the outflow velocity as function of time (pulsation), a velocity gradient across the launching
area (shear) and a small opening angle at the outflow footpoint (spray). Our outflow sub-grid model also includes shear and spray as free parameters,
but the precession and velocity variations of our outflows are determined self-consistently by the properties of our sink particles.

As in \citet{voelker99}, the H$_2$ emission comes predominantly from the outflow head. However, we also see significant emission from the
bulk of the outflow because the amplitude of our velocity pulsations is not limited, while \citet{voelker99} do not allow the outflow
velocity to grow by more than a factor of two during a pulsation period. We have therefore much more shock-heated gas in the interior
of our outflow. The \citet{voelker99} Hammer model leads to regularly-spaced mini-bow shocks that are aligned
with the bulk flow.

Our outflows look considerably more irregular for a number of
reasons. First, our precession angles grow to markedly larger values
than the few degrees considered by \citet{voelker99} (see
Figure~\ref{fig:anglepsi}). Second, the momentum output from each star can vary by
more than an order of magnitude (see
Figure~\ref{fig:momentum}). Third, we have four outflows interacting
with each other as they propagate instead of just a single outflow
expanding in a quiescent medium.

\begin{figure*}[t]
\begin{center}
\includegraphics[width=400pt]{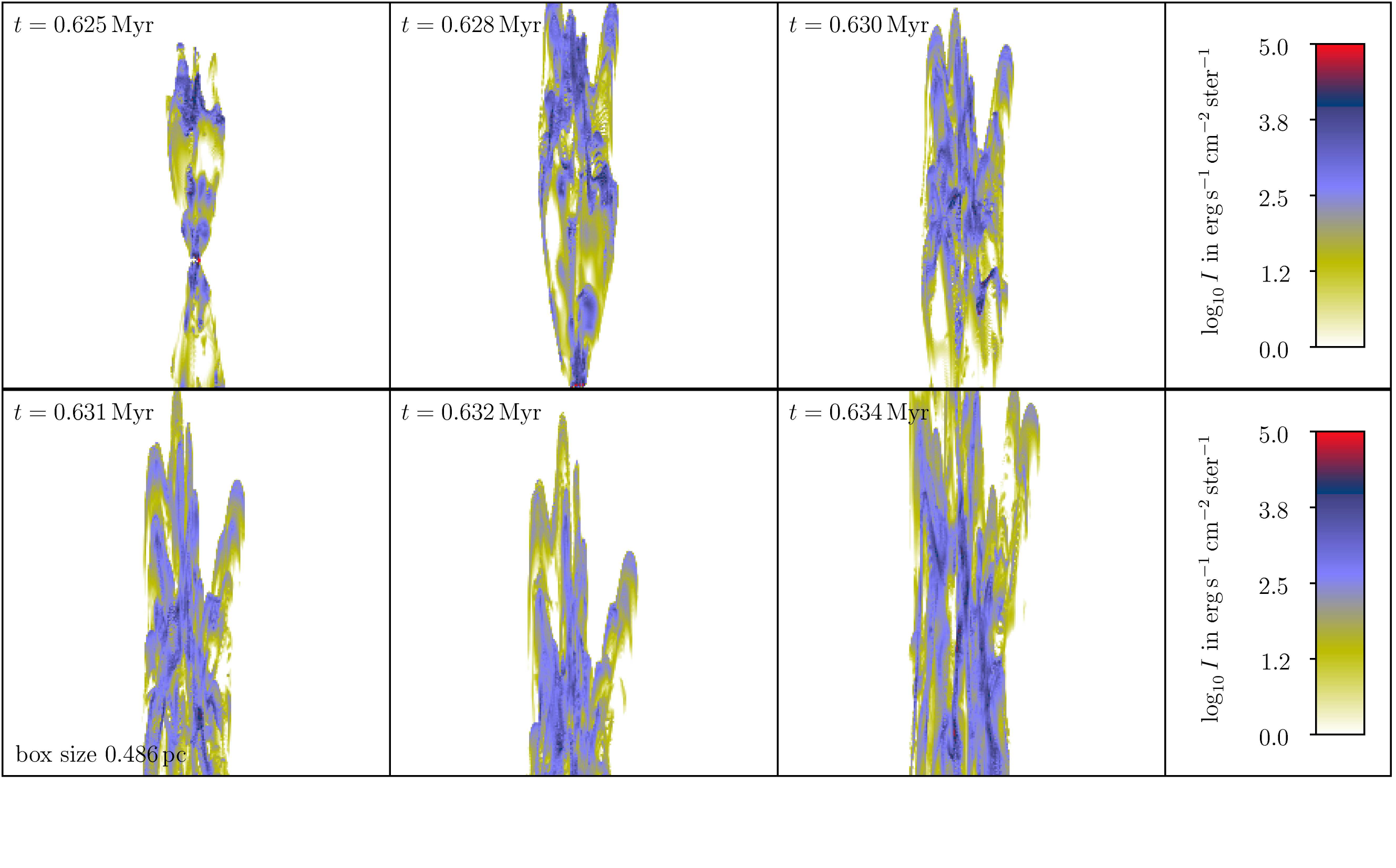}
\end{center}
\vspace{-20pt}
\caption{Intensity of the H$_2$ S(1) 1-0 line for the six snapshots of Figure~\ref{fig:head}. The spatial scale is the same in all four images.}
\label{fig:headh2}
\end{figure*}

In Figure~\ref{fig:headh2} we show the outflow heads from Figure~\ref{fig:head} in H$_2$ emission. The head structure with its multiple
tips looks very similar to the density slice in general. Shocks running through the outflow can
be identified through their stronger emission, because the shocks are generally denser than the outflow interior.
The brightness of the emission at the outflow tip depends on when a shock has reached this tip for the last time. This is because the shock
rapidly heats up the gas, which then subsequently cools. The H$_2$ emission then diminishes with the gas temperature.
This effect can be nicely seen in the animated version of Figure~\ref{fig:panelplotH2}.

\begin{figure*}[t]
\centerline{\includegraphics[width=340pt]{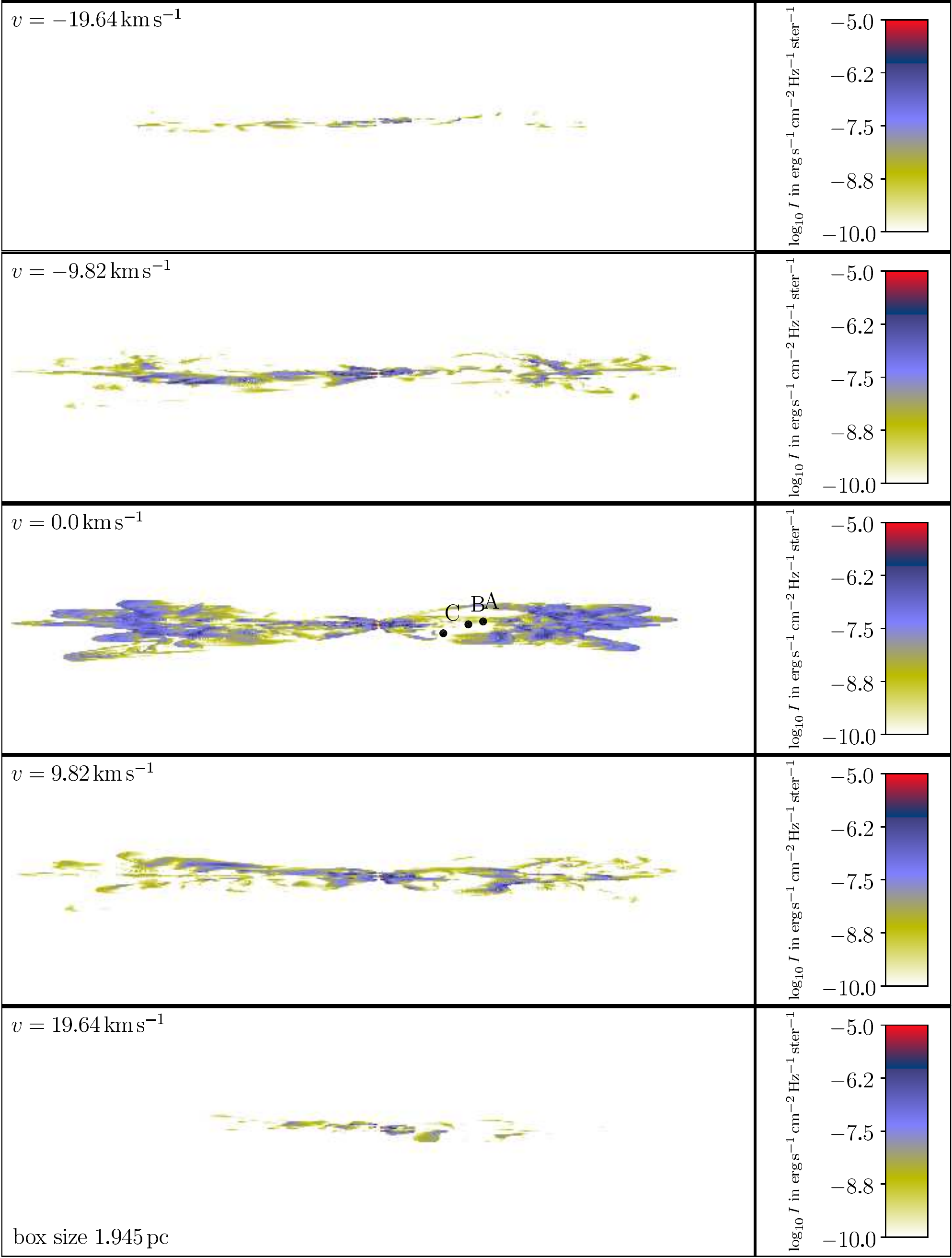}}
\caption{Edge-on channel maps of the H$_2$ S(1) 1-0 line for the
  collective outflow at $t = 0.630\,$Myr. Spectra through the marked
  points are shown in Figure~\ref{fig:spectra1}.} 
\label{fig:h2linech1}
\end{figure*}

\begin{figure*}[t]
\centerline{\includegraphics[width=340pt]{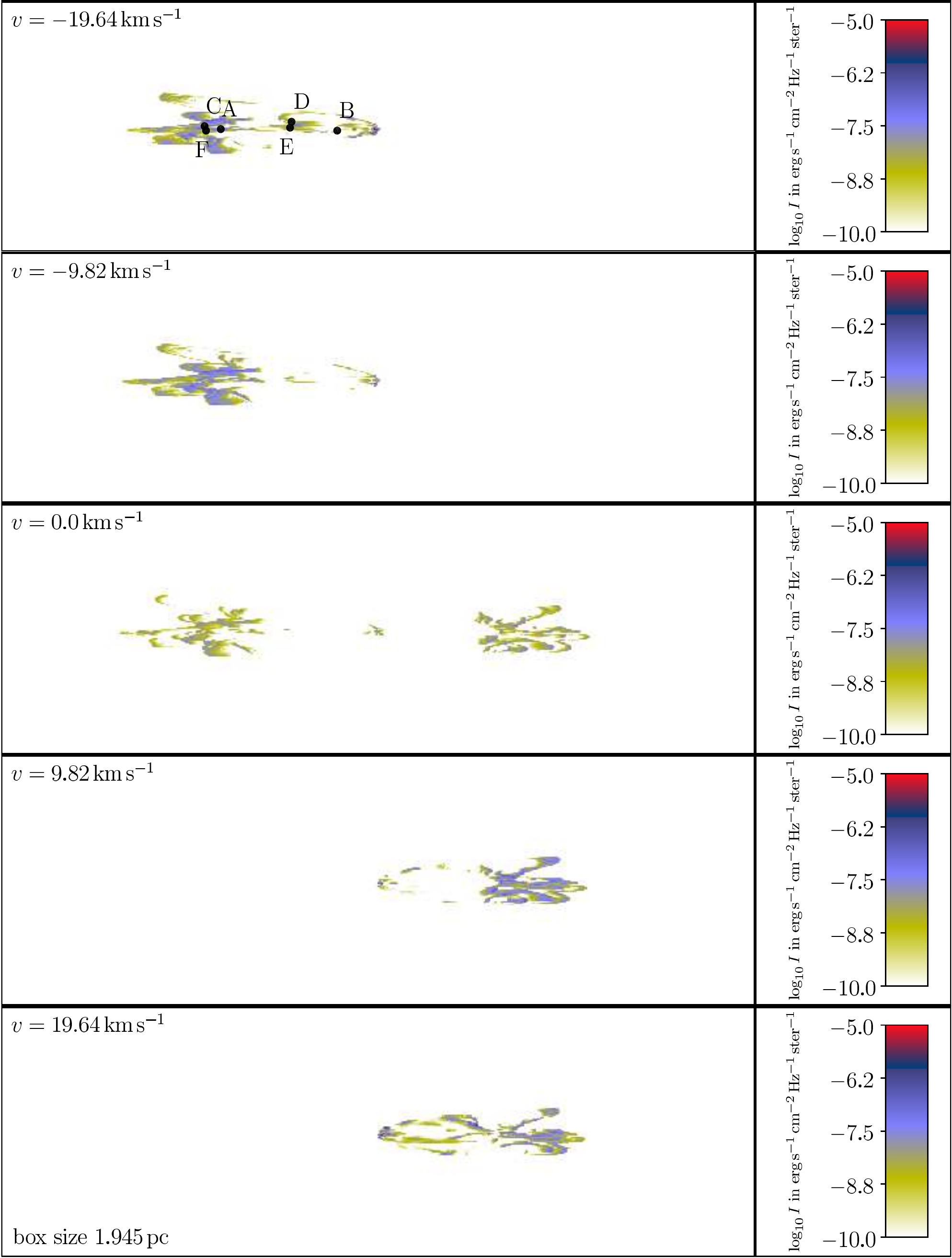}}
\caption{Channel maps at an inclination of 45$^\circ$ of the H$_2$ S(1) 1-0 line for the collective outflow at $t = 0.630\,$Myr. Spectra through the marked points are shown in Figure~\ref{fig:spectra}.}
\label{fig:h2linech2}
\end{figure*}

We study the kinematics for a simulation snapshot at $t = 0.630\,$Myr,
when the outflow is still completely contained inside the simulation
box. 
To compare the kinematics of the collective outflow to observations, we have generated maps at an inclination of 45$^\circ$ and 90$^\circ$
at a spatial resolution of 191~AU and a spectral resolution of 9.82~km~s$^{-1}$ (63 channels). This corresponds to the observing conditions of \citet{hiriart04}.
We show channel maps of the observed outflow in Figure~\ref{fig:h2linech1} and \ref{fig:h2linech2} for the two inclination angles. As expected,
in the edge-on maps the emission is strongest near the rest velocity, while the 45$^\circ$ maps have larger emission at finite velocities.
The finger-like structure of the outflow heads is visible in almost all channels.

\begin{figure*}[t]
\centerline{\includegraphics[width=340pt]{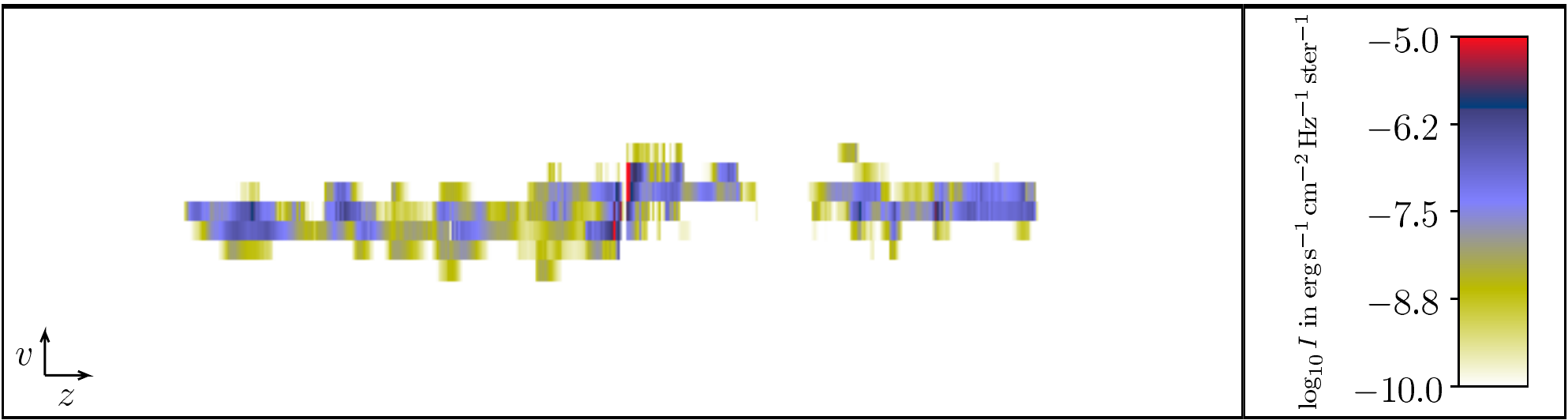}}
\caption{PV slice through the horizontal midline of Figure~\ref{fig:h2linech1}. The scaling in the horizontal direction is identical to Figure~\ref{fig:h2linech1}.
The vertical axis extends from -98.2~km~s$^{-1}$ to +98.2~km~s$^{-1}$ in steps of 9.82~km~s$^{-1}$.} 
\label{fig:h2linepvslice1}
\end{figure*}

\begin{figure*}[t]
\centerline{\includegraphics[width=340pt]{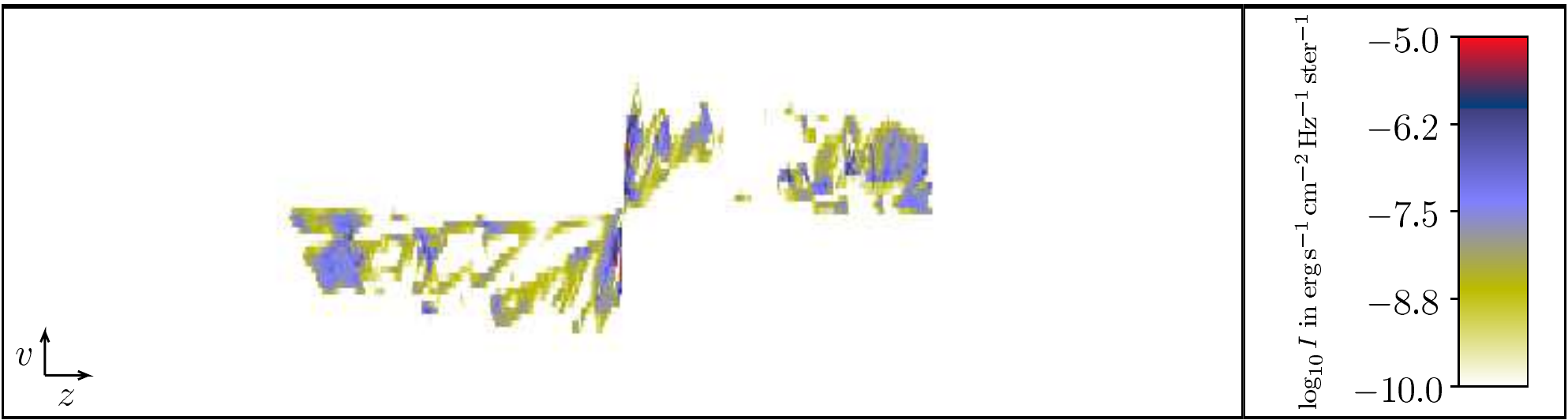}}
\caption{PV slice through the horizontal midline of Figure~\ref{fig:h2linech2}. The scaling in the horizontal direction is identical to Figure~\ref{fig:h2linech2}.
The vertical axis extends from -304~km~s$^{-1}$ to +304~km~s$^{-1}$ in steps of 9.82~km~s$^{-1}$.} 
\label{fig:h2linepvslice}
\end{figure*}

We present a position-velocity (PV) diagram for the 90$^\circ$ map in
Figure~\ref{fig:h2linepvslice1} and for the 45$^\circ$ map in
Figure~\ref{fig:h2linepvslice}. The slices were taken perpendicular to
the disk through the center of the group, corresponding to the
horizontal midline in the Figure~\ref{fig:h2linech1} and
\ref{fig:h2linech2} channel maps. The emission covers the PV plane
only fragmentarily, which may hint at the different components of the
collective outflow. However, it is impossible to identify the four
individual outflows because of the strong interaction between
them. The radial spikes visible in the 45$^\circ$ PV diagrams are
typical of pulsation and density enhancements in the outflow
\citep{smith97,voelker99}, in our case driven by the variations in
accretion rate.

\begin{figure*}[t]
\centerline{\includegraphics[width=160pt]{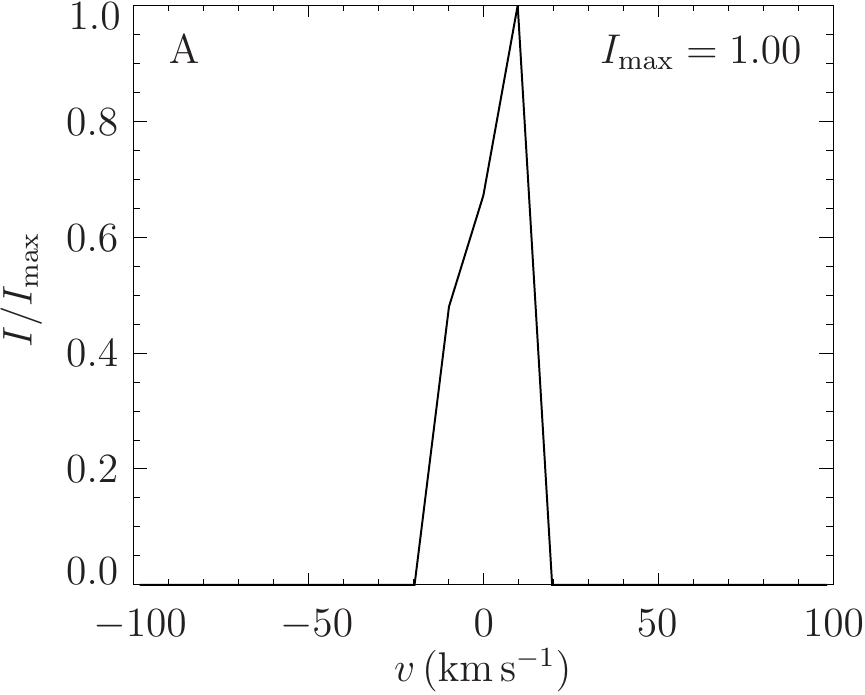}
\includegraphics[width=160pt]{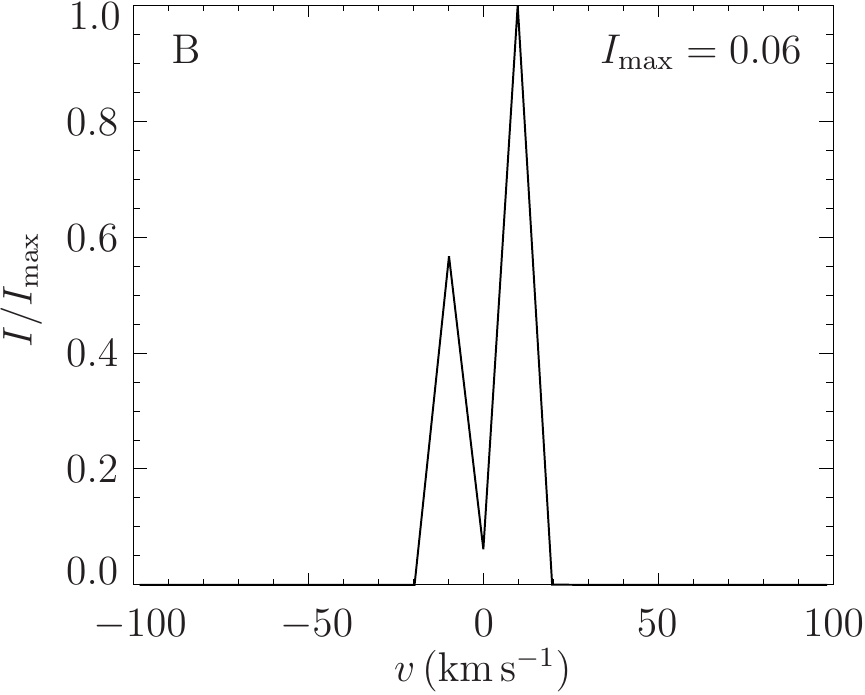}
\includegraphics[width=160pt]{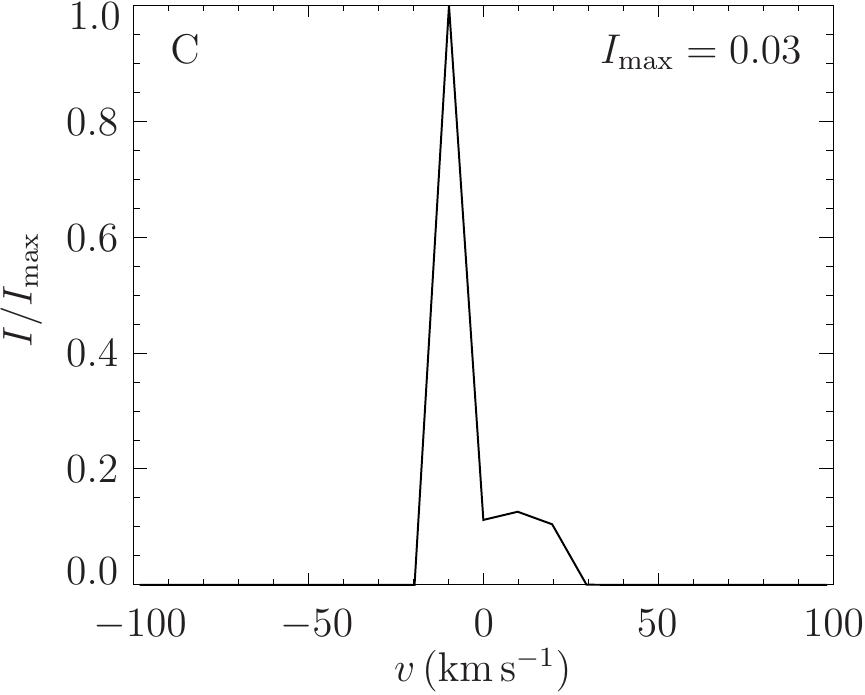}}
\caption{Spectra taken through the three positions in the
  outflow viewed from 90$^{\circ}$ indicated in
  Figure~\ref{fig:h2linech1}. The position of the spectra and the
  relative maximum intensities are indicated.}
\label{fig:spectra1}
\end{figure*}

\begin{figure*}[t]
\centerline{\includegraphics[width=160pt]{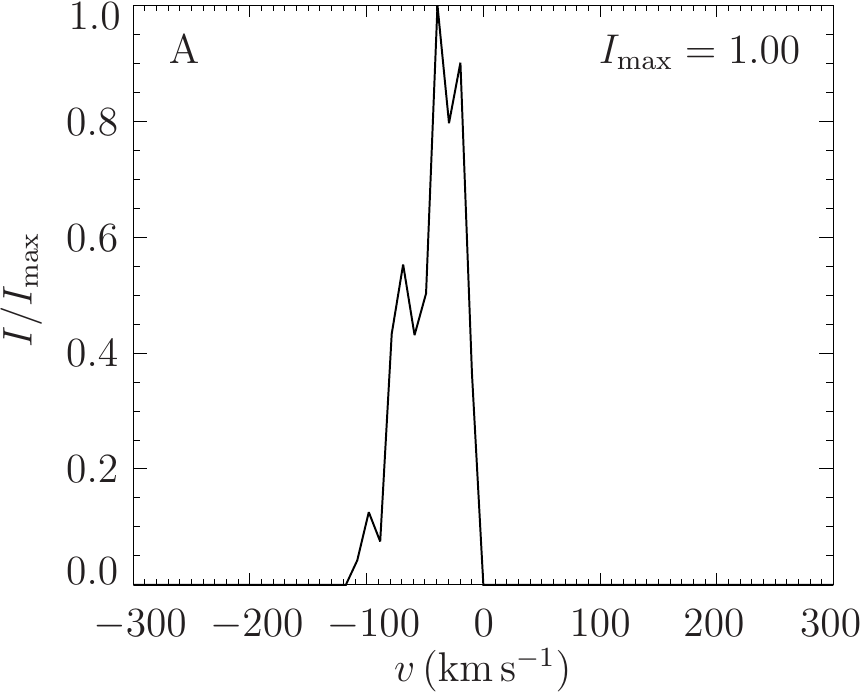}
\includegraphics[width=160pt]{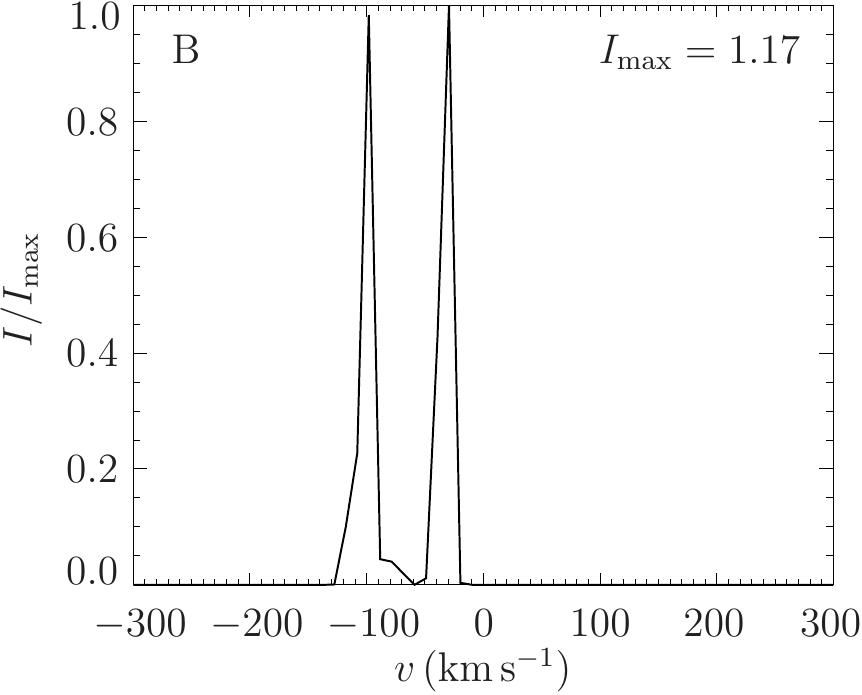}
\includegraphics[width=160pt]{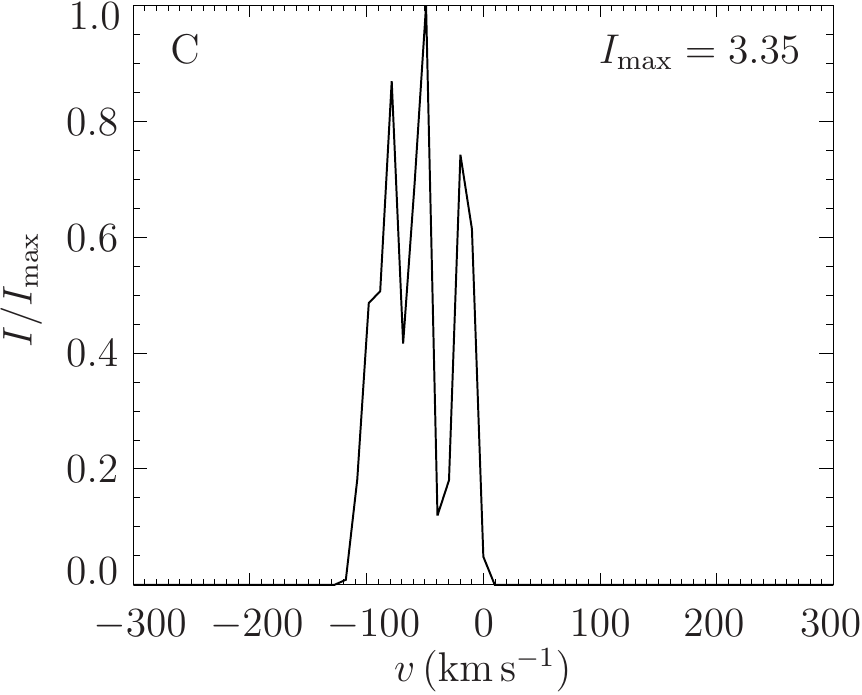}}
\centerline{\includegraphics[width=160pt]{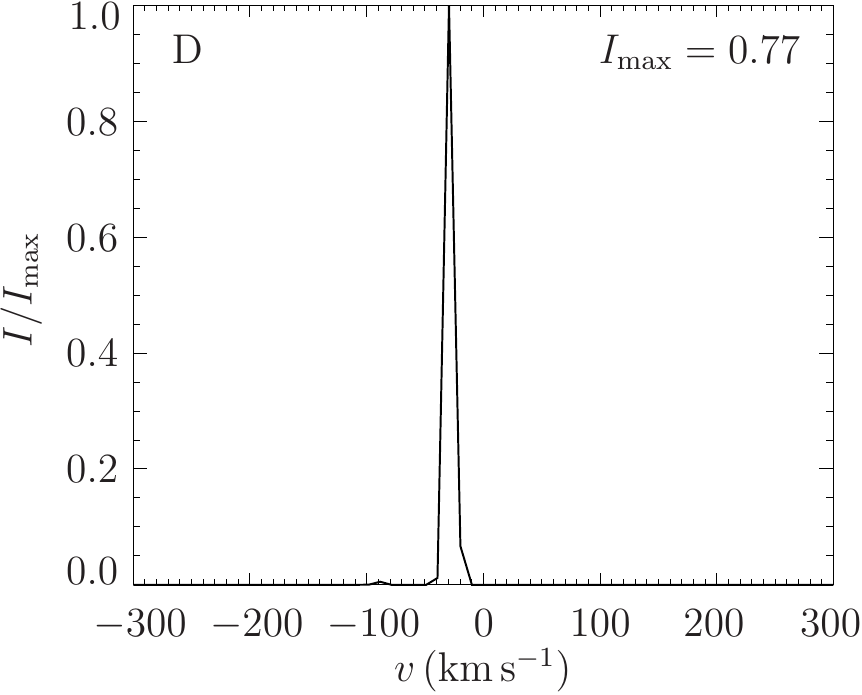}
\includegraphics[width=160pt]{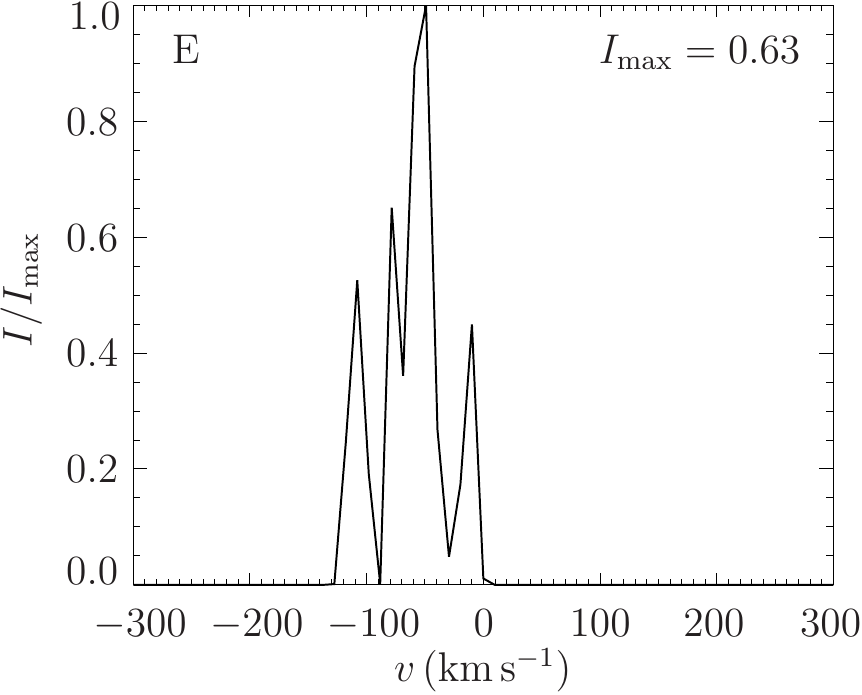}
\includegraphics[width=160pt]{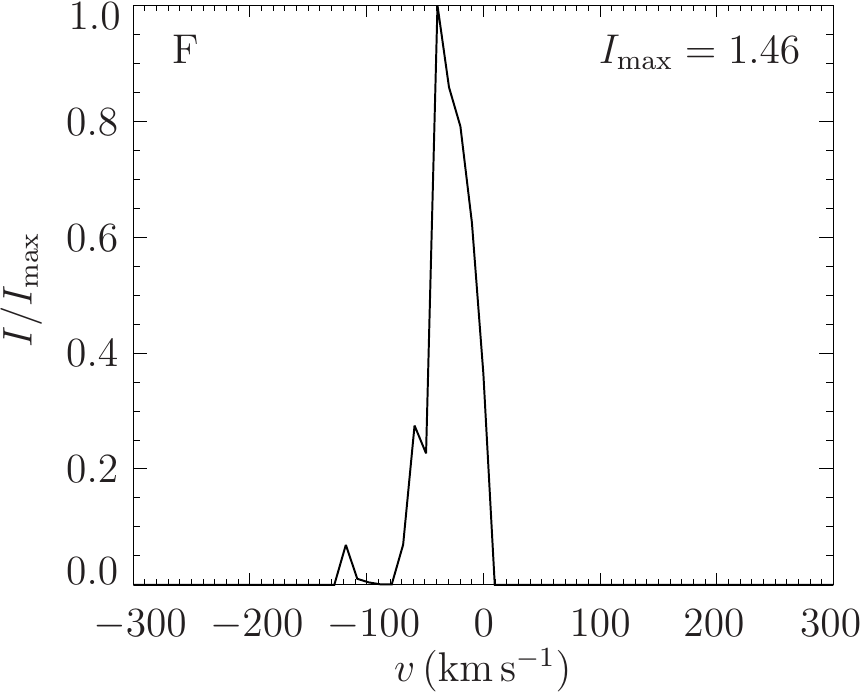}}
\caption{Spectra taken through six positions in the
  outflow viewed from 45$^\circ$ indicated in
  Figure~\ref{fig:h2linech2}. The position of the spectra and the
  relative maximum intensities are indicated.}
\label{fig:spectra}
\end{figure*}

Spectra for positions marked in the PV diagrams for the 90$^{\circ}$
view 
   (in 
Figure~\ref{fig:h2linech1}) and the 45$^{\circ}$ view 
   (in
Figure~\ref{fig:h2linech2}) are displayed in Figure~\ref{fig:spectra1}
and Figure~\ref{fig:spectra}, respectively. The observation at 45$^\circ$ shows
more complex spectra since the line-of-sight velocities are much
greater in this case. We find a variety of spectral forms, ranging
from a sharp single peak through broad single peaks to multiply-peaked
spectral profiles. The spectrum is a strong function of position, and
there is no apparent systematic structure, except that the spectrum
becomes more complex in regions where shock fronts overlap, which
generally correponds to the brightest regions in
Figure~\ref{fig:h2linech2}.

\subsubsection{CO Observations of the Collective Outflow}
\label{sec:synobsco}

We model emission from the $J=2-1$ transition of CO assuming local thermodynamic equilibrium (LTE), with
the Einstein coefficient taken from the Leiden Atomic and Molecular
Database \citep{schoeieretal05}, and a constant 
ratio of H$_2$ to CO abundance [H$_2$]/[CO]=$1\times10^{-4}$.
ALMA observation simulations were carried out using CASA \citep{mcmullin07}, version 4.1. 
The intensities of the output image cubes from RADMC-3D were converted
to Jansky units by assuming a
source distance of 3~kpc.  The source coordinates were chosen for comparison to a southern hemisphere
analog of DR21 (with a declination of -42$^\circ$), however all coordinates in the resulting figures
are given in spatial offsets from the phase center. A southern declination was necessary for properly
sampling the UV plane.

We simulated observing CO ($J=2-1$) at a spectral and spatial
resolution of 0.5~km~s$^{-1}$ and 0.22$''$ using one of the full ALMA
configurations bundled into CASA (alma.out08.cfg). The simulations
were run assuming four hours of integration, and a precipitable water
vapour column of 1\,mm. This is consistent with good weather observing
for Band 6 (230 GHz) ALMA observations, allowing an accurate
representation of atmospheric noise in the visibilities.  The online
sensitivity calculator estimates a root-mean-square noise level of
4~mJy~beam$^{-1}$, and we used a threshold slightly higher
than this (6~mJy~beam$^{-1}$) as our 1$\sigma$-level when cleaning.
Cleaning was done non-interactively, without the use of clean boxes.
We used natural weighting, and our final image cubes spanned
384$\times$384$\times$256, where the last dimension indicates the
number of spectral channels.

\begin{figure*}[t]
\centerline{\includegraphics[height=350pt]{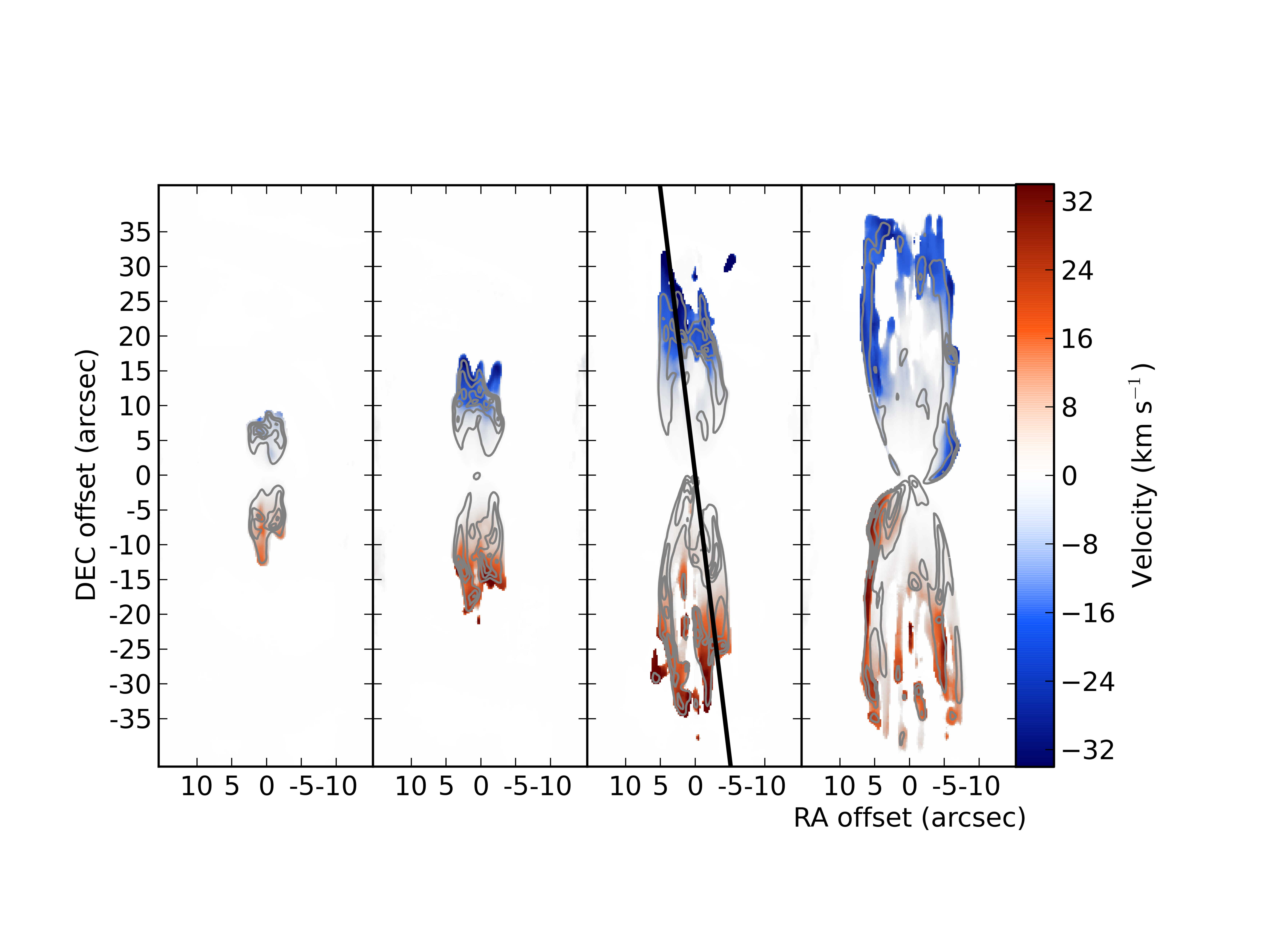}}
\caption{First moment maps of the CO $J=2-1$ transition for the four snapshots from Figure~\ref{fig:panelplot} at an inclination of 30$^\circ$.
The four contours in each plot are linearly spaced between 20 and 90\% of the peak intensity in the zeroth moment (integrated intensity) map. The black line
represents the PV slice shown in Figure~\ref{fig:copv}.}
\label{fig:co1stmom}
\end{figure*}

We show the first moment (intensity-weighted velocity) maps of the snapshots from Figure~\ref{fig:panelplot}
at an inclination of 30$^\circ$ in Figure~\ref{fig:co1stmom}.
Note that the size of the mapped region grows with the outflow scale. Also notice
that the outflow has already started leaving the simulation box in the third snapshot, and that a significant portion of it is
missing in the fourth. The inferred mass and kinematics of the
collective outflow in these cases is, of course, only a lower
bound on the true values.

The CO maps prominently exhibit the outflow heads with multiple tips in each of them. The maximum outflow velocity grows with time,
but since the maximum value in the first moment maps is attained in the heads, this trend seems to break with the fourth
snapshot. In actual fact, the outflow velocity is nearly constant along the outflow. It is the increased density near the
outflow boundary that, via the intensity-weighting in the first moment maps, leads to the impression that the velocity
in the head was higher. Note that no property of the outflow
morphology suggests that the outflow might be driven by
several protostars. Although the head does have multiple tips, they
come and go in a way consistent with Rayleigh-Taylor instability (see
Sect.~\ref{sec:outevo}) 

A PV slice through the third outflow from Figure~\ref{fig:co1stmom} is shown in Figure~\ref{fig:copv}. One can distinguish
multiple components, but again it is not possible to clearly associate them with the four protostars. The PV slice
shows that the emission from the collective outflow can be decomposed into three parts: 
(1) small-scale low-velocity gas, (2) large-scale high-velocity gas, and (3) large-scale medium-velocity gas.
Component (1) is produced by the disk-like structure that contains the four protostars.
Components (2) and (3) trace different pieces of the outflow, which are possibly produced by different protostars.
The CO bullets correspond to shocks running through the outflow and are also visible in Figure~\ref{fig:co1stmom}.

The image cubes for each snapshot were analyzed to determine the outflow mass, momentum, and energy. 
Using a kinematic age for the outflow, we also determined an outflow mechanical luminosity and mass loss rate for each snapshot.
The total outflow mass was determined by summing the mass in each velocity bin outside of the central 2~km~s$^{-1}$.
The gas mass in an individual velocity bin was determined by multiplying the average intensity by a multiplication factor
based on the ambient temperature (in this case 30~K). 
This results in the average number of emitting particles per unit area, assuming that the emission is optically thin.
This average number of emitting molecules was then multiplied by the emitting area to determine the total number of
emitting molecules. This was then further multiplied by the abundance ratio for CO, and the mean molecular weight $2.3$.
More details on this type of calculation can be found in
\citet{klaawils07}.
The kinematic age of each snapshot was calculated based on the extent of the outflow and the intensity-weighted velocity at
the edge of the outflow (12,~30,~36, and~31~km~s$^{-1}$ respectively for the earliest to latest snapshot). The timescale was then
determined by dividing the distance travelled by the velocity at which the gas is moving. These kinematic ages allowed
for the outflow mechanical luminosity and mass loss rate to be determined.
To calculate the outflow momentum and energy, the mass in each bin was
multiplied by the velocity of that bin, and the momenta
($P=M\times V$) and energy ($E=0.5M\times V^2$) were summed over all
channels with velocities greater than 2~km~s$^{-1}$.

\section{Observational Comparisons}

We report results from early times in the development of the
collective outflow from a massive star forming region. In particular,
none of the protostars in our model has yet reached a high enough mass, and
ionizing luminosity, to form an ultracompact H~{\sc ii} region. There
are no regions passing through this short-lived phase with distances less
than a few kiloparsecs, close enough for their jets to be clearly
resolved and imaged. However, at least two rather closer regions, Cepheus~A
and DR~21, appear to be in the immediately subsequent phase, with a major
outflow associated with small, highly obscured ultracompact H~{\sc ii}
regions. We therefore focus our comparison of our results to these two regions,
after beginning with a general discussion of the outflow kinematics.

\subsection{Kinematics of the Collective Outflow}

All of the 
  outflow properties derived from our simulated CO observations 
are shown in Table~\ref{table:co}. The kinematic ages
on which the calculation of the luminosity and mass-loss rates is based are also shown.
The second snapshot appears younger than the first one because its intensity-weighted velocity grows
disproportionally to its size. This can be explained by the fact that shortly before snapshot two, the third
and fourth star in the group have formed (compare Figure~\ref{fig:accrhist}), which already contribute to the collective
outflow at its base but have not influenced its spatial extent yet.
Table~\ref{table:co} also contains the real outflow ages, measured from the formation of the first
sink particle.
All measured quantities---the mass, momentum, kinetic energy, mechanical luminosity, and mass-loss rate---match
well with the observed properties of high-mass outflows \citep[e.g.][]{beutheretal02,wuetal04}.

We can compare the outflow kinematics measured from the simulated CO
emission map to the actual model values.
We determine the actual values of the mass, momentum and kinetic energy by summing over all cells
above the disk plane with positive vertical velocity components, and
similarly below the disk plane with negative vertical velocities. The resulting
values are shown in Table~\ref{table:sim}. All masses and energies agree within a factor of two with the
observationally determined numbers, while for the momentum the
difference can be as much as a factor of three.
This level of agreement is consistent with the results of a similar study for outflows in magnetohydrodynamical
simulations \citep{petersetal14}.
The kinematic and real ages also agree within a factor of two.

We performed $^{13}$CO and C$^{18}$O ALMA simulations for two of the snapshots to determine
the optical depth of the simulated observations of $^{12}$CO. We found that the emission is not very opaque.
Accounting for the opacity of the $^{12}$CO line changed our derived outflow properties by less than a few percent.

We can compare these results with high resolution observations of the early stages of high-mass star formation. 
We note that statistical studies of these regions are not at high resolution, and further high resolution
observations are required for this type of analysis.

The energetics in the collective outflow are slightly lower than those observed in the group of outflows
in 20293+3952 observed by \citet{beuschgue04}, although the masses are close to those observed in this source,
suggesting the velocities of the gas in the simulation are lower than those observed by \citet{beuschgue04}.
Conversely, we find masses higher than those observed by \citet{palau13} towards 19520+2759 MM1, but
energies similar to their values and with similar dynamical timescales.

\begin{figure}[t]
\centerline{\includegraphics[height=180pt]{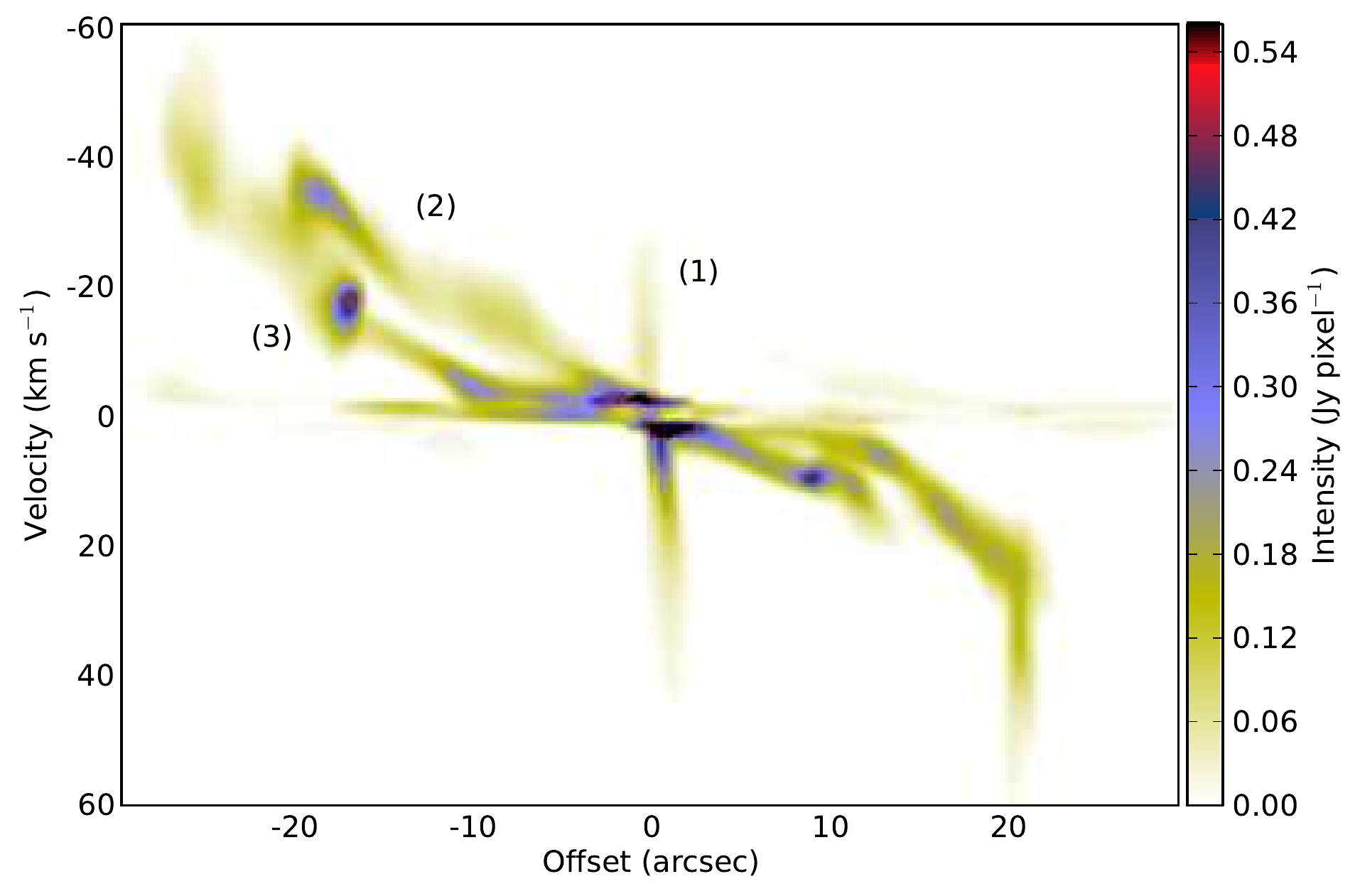}}
\caption{PV slice in CO $J=2-1$ for the third snapshot from Figure~\ref{fig:co1stmom} at a 7$^\circ$ offset from the vertical axis.
The cut goes through the furthest blue peak in Figure~\ref{fig:co1stmom}. The plots shows three different components:
(1) small-scale low-velocity gas, (2) large-scale high-velocity gas, and (3) large-scale medium-velocity gas.}
\label{fig:copv}
\end{figure}

The mass, momenta and energetics are  all 2--3 orders of magnitude larger than those seen by \citet{arce13}
in HH 46/47, which is expected since the stars in our simulations are
substantially more massive than those powering HH 46/47. 
What is interesting is that in Figure \ref{fig:copv}, we see multiple outflow components, just as they do
(compare their Figure 9).

\citet{qiu09} calculated the outflow energetics for G240.31+0.07 and all of their values are approximately
a factor of 5 greater than those determined here. Their region already shows evidence for an \hii\ region, and may
be in a later evolutionary stage than the region being studied here. Similarly, \citet{zap09} show the
collimated outflow from a cluster of forming stars in W51N at a similar evolutionary stage to that of the \citet{qiu09} study.

These comparisons show that our outflow has higher masses and energetics than low-mass regions studied at similar
resolutions, and lower values than those expected for more evolved clusters of forming stars: those that have already
formed their \hii\ regions. 

\begin{table*}
\caption{Outflow properties determined from synthetic CO observations}
\label{table:co}
\begin{center}
\begin{tabular}{lrcccccc}
\hline
&& \multicolumn{1}{c}{$M$} & \multicolumn{1}{c}{$P$} & \multicolumn{1}{c}{$E$} & \multicolumn{1}{c}{$L$} & \multicolumn{1}{c}{$\dot{M}$}
& \multicolumn{1}{c}{$t$}\\
&& \multicolumn{1}{c}{($M_\odot$)} & \multicolumn{1}{c}{($M_\odot$ km s$^{-1}$)} & \multicolumn{1}{c}{(10$^{44}$ erg)}
& \multicolumn{1}{c}{($L_\odot$)} & \multicolumn{1}{c}{($10^{-3}\,M_\odot\,$yr$^{-1}$)}
& \multicolumn{1}{c}{(kyr)}\\ \hline
$t = 0.624$\,Myr &
blue& 13.0 & 136 & 179 & 12.4 & 1.09 & 12.0 \\
&red& 11.7 & 147 & 232 & 16.1 & 0.99 &      \\
\hline
$t = 0.628$\,Myr &
blue& 23.9 & 352 & 716 & 77.6 & 3.15 & 7.6 \\
&red& 18.7 & 303 & 672 & 72.9 & 2.46 &     \\
\hline
$t = 0.635$\,Myr &
blue& 30.4 & 487 & 1080 & 72.5 & 2.48 & 12.3 \\
&red& 22.8 & 380 & 866 & 58.1 & 1.86 &       \\
\hline
$t = 0.642$\,Myr &
blue& 16.4 & 224 & 438 & 20.7 & 0.94 & 17.5 \\
&red& 23.2 & 391 & 893 & 42.2 & 1.33 &      \\
\hline
\end{tabular}
\medskip\\
Outflow mass $M$, momentum $P$, kinetic energy $E$, mechanical luminosity $L$, mass-loss rate $\dot{M}$
and outflow age $t$ as determined from the synthetic CO observations.
\end{center}
\end{table*}

\begin{table*}
\caption{Outflow properties determined from the simulation data}
\label{table:sim}
\begin{center}
\begin{tabular}{lrcccccc}
\hline
&& \multicolumn{1}{c}{$M$} & \multicolumn{1}{c}{$P$} & \multicolumn{1}{c}{$E$} & \multicolumn{1}{c}{$L$} & \multicolumn{1}{c}{$\dot{M}$}
& \multicolumn{1}{c}{$t$}\\
&& \multicolumn{1}{c}{($M_\odot$)} & \multicolumn{1}{c}{($M_\odot$ km s$^{-1}$)} & \multicolumn{1}{c}{(10$^{44}$ erg)}
& \multicolumn{1}{c}{($L_\odot$)} & \multicolumn{1}{c}{($10^{-3}\,M_\odot\,$yr$^{-1}$)}
& \multicolumn{1}{c}{(kyr)}\\ \hline
$t = 0.624$\,Myr &
blue& 15.3 & 45.5 & 252 & 20.7 & 1.53 & 10.0 \\
&red& 14.4 & 48.4 & 338 & 27.8 & 1.44 &      \\
\hline
$t = 0.628$\,Myr &
blue& 16.9 & 105 & 745 & 43.8 & 1.21 & 14.0 \\
&red& 12.0 & 109 & 893 & 52.5 & 0.86 &      \\
\hline
$t = 0.635$\,Myr &
blue& 19.8 & 216 & 2230 & 90.5 & 0.98 & 20.3 \\
&red& 14.0 & 211 & 2260 & 91.7 & 0.69 &      \\
\hline
$t = 0.642$\,Myr &
blue& 21.9 & 230 & 1660 & 50.1 & 0.80 & 27.3 \\
&red& 14.5 & 198 & 2050 & 61.7 & 0.53 &      \\
\hline
\end{tabular}
\medskip\\
Outflow mass $M$, momentum $P$, kinetic energy $E$, mechanical luminosity $L$, mass-loss rate $\dot{M}$
and outflow age $t$ as measured from the simulation data.
\end{center}
\end{table*}

\subsection{Cepheus A}

We now turn to a detailed comparison with nearby objects.  Cepheus A
is a massive star-forming region lying at a distance of 700 +30/-28 pc,
based on Very Long Baseline Array (VLBA) parallax measurements to two of the radio continuum
sources it contains \citep{moscadelli09,dzib11}.
This is consistent with previous estimates of at least 620~pc
\citep{de-zeeuw1999}, and possibly as far as 730~pc
\citep{johnson1957,crawford1970}. 

Cep A contains young stellar objects with a total bolometric
luminosity of $2.5 \times 10^4$~L$_{\odot}$, exceeding that of a
main-sequence B0 star \citep{evans1981,ellis1990}, lying behind a
visual extinction of 90 magnitudes. \citet{hughes1984} detected
multiple radio sources within the region, now denoted HW~1--9,
consistent with 14 B3 stars with masses of around 8~M$_{\odot}$ lying
within a 0.1 pc radius.  A rotating, dense core has been traced in CS
emission with a radius of about 0.16~pc and a dynamical mass of
330~M$_{\odot}$ \citep{narayanan1996}. The brightest source, HW~2, has
about half the total luminosity, suggesting it is a B0.5 star
approaching 15~M$_{\odot}$ \citep{hughes1995}. A Keplerian disk a
hundred times smaller was resolved around HW~2 using CH$_3$CN line
emission mapped with the Submillimeter Array \citep{patel2005}. A jet
whose knots have proper motions of order 500 km~s$^{-1}$ has been
observed in radio continuum within HW~2 \citep{curiel2006}.  Maser
activity reveals at least two additional centers of star formation
activity within 200 AU of HW~2 \citep{torrelles2001,gallimore2003}.
A cluster of class~I and class~II YSOs also lies in this
location \citep{gutermuth2005}. The surrounding gas core traced by
CS has a dynamical mass of 330~M$_\odot$ within a radius of 0.32~pc
\citep{narayanan1996}. \citet{rodriguez1980} initially
identified the major east-west bipolar CO outflow associated with
Cep~A.  Additional components are aligned northeast-southwest \citep{bally1991}.
CO radial velocities as high as 70 km~s$^{-1}$ have
been observed in the central 2' \citep{narayanan1996}, suggesting
a powerful outflow has only recently become active.

\citet{hiriart04} resolved the emission from the H$_2\, v = 1-0$ S(1)
emission line with the spectral parameters used in our Figures~\ref{fig:h2linech1}
and~\ref{fig:h2linech2}.  Their Figure~1 shows the same characteristic clumpiness over
the same velocity scale as seen in those Figures, including a hint of
the highly structured leading bow shock seen in our models. Our
edge-on Figure~\ref{fig:h2linech1} is the better match to the observations, as Cep A
lies nearly in the plane of the sky. This morphology is much better
resolved in the line image shown in Figure~1 of \citet{cunningham2009}.
This clearly shows the same multiple structures
resembling bow shocks in both the interior and leading edge of the jet
that are seen in our Figures~\ref{fig:panelplotH2} and~\ref{fig:headh2}. \citet{cunningham2009} trace
linear features through these structures and attribute them to a
single, precessing object with arbitrarily chosen parameters, but our models self-consistently reproduce
the same morphology with interacting, time-variable jets from the
multiple sources forming during gravitational collapse of a core.
\citet{hiriart04} also show a position velocity diagram in their
Figure~4 generally consistent with our Figure~\ref{fig:h2linepvslice}.  In particular, we
reproduce the characteristic variation of the peak velocity along the
length of the jet, due to intersections with interior shocks having
high transverse velocity components.

\subsection{DR 21}

The distance to the next nearest massive star-forming region, DR~21 in
the Cygnus X GMC complex \citep{schneider2006} has been determined to
be 1.4~kpc by VLBA maser parallax measurements
\citep{rygletal12}, confirming the 1.5~kpc estimate of
\citep{odenwald1993}.
Like Cepheus A, DR~21
contains a massive core with multiple ultracompact and hypercompact
H~{\sc ii} regions \citep{harris1973,dickel1986,cyganowski2003}
linked to ZAMS O stars \citep{roelfsema1989}; water, OH,
and methanol masers \citep{plambeck1990,araya2009}; many
embedded IR sources; and a complex molecular outflow nearly in the
plane of the sky \citep{davis2007}. \citet{zapata2013} showed
multiple outflows traced by CO, SiO and methanol associated with the
DR~21(OH) hot core roughly 2~pc away from the DR~21 H~{\sc ii} region,
and probably part of the same star formation event.

High-resolution images of DR~21 with a narrow-band H$_2\,v = 1-0$ S(1)
emission line filter by \citet{davis1996}, \citet{cruzgonzalez2007},
and \citet{davis2007} revealed morphology strongly
reminiscent of that described above for Cep A, including multiple
interior structures resembling bow shocks, and a complex, structured
jet head, as shown in Figure~2 of \citet{davis2007}. 

\citet{davis1996} measured line profiles with 20~km~s$^{-1}$ velocity
resolution, finding both asymmetric, blue-shifted profiles, and also
more symmetric profiles (see their Figures~6 and~7). They often
observed multicomponent structure in both sorts of profiles. In our
Figure~\ref{fig:spectra1}, we show line profiles from our run viewed
at 90$^{\circ}$ (edge on) with 9.82~km~s$^{-1}$ resolution. These
include a clearly blue-shifted profile with multiple components, and
symmetric profiles with and without multiple components. This agrees
well with the somewhat lower spectral resolution \citet{davis1996}
results. \citet{cruzgonzalez2007} published a position-velocity
diagram (their Figure 6), again showing the characteristic variation
in the velocity direction associated with clumpiness in the emission
seen in our Figure~\ref{fig:h2linepvslice}. Our model does not
reproduce the large scale structure seen by \citet{cruzgonzalez2007},
though, perhaps because of our choice of smooth initial conditions,
which reduces the amount of large-scale density structure that the jet
interacts with.

\subsection{Summary}

We have demonstrated that an outflow driven by multiple stars formed
by fragmentation of the accretion flow onto a massive star appears
consistent with observations of massive stellar outflows.  We have
done this in two ways. First, we compared the integrated mass and
momentum of our simulated outflow to typical examples of massive
stellar outflows at similar early stages in their evolution, finding
that the simulated values lie within the observed ranges.
We note that other mechanisms we have examined, such as magnetic
driving \citep{petersetal11a} or ionization \citep{petersetal12b}, fail
this fundamental test.

Second, we compared our models to the relatively well resolved
outflows from Cep~A and DR~21. These share several characteristics. They
have a chaotic structure with multiple apparent bowshocks, and
relatively poor collimation. In particular, they do not resemble the
well-defined structure of isolated, low-mass stellar jets,
with a train of Herbig-Haro objects ending in a single bow shock,
often connected by a well-defined high velocity beam. We argue that
these multiple bow shocks can be naturally explained by
Rayleigh-Taylor instability acting on the broad head of the jet as it
propagates down the density gradient of the envelope; while the poor
collimation occurs as the angular momenta of the sources gradually
diverge from a common direction.

The kinematics of the massive outflows confirm these morphological
impressions, with multiple velocity components leading some workers to
characterize the flow as turbulent
\citep{davis1996,hiriart04,cruzgonzalez2007}.  Despite these chaotic
tendencies, the jets do show general collimation, suggesting
origination from stars formed from material sharing a common angular
momentum axis, as would be expected in a set of stars formed from any
single, gravitationally collapsing region, even in the presence of a
turbulent background flow. All of these morphological
and kinematic characteristics are also seen in the combined outflow
generated by multiple, time-variable jets in our model of massive star
formation.

\section*{Acknowledgements}

We thank R.~Banerjee for contributions to the development
of the outflow subgrid model. We further acknowledge useful discussions with
J. Bally, J. Mackey, \& M. Krumholz, as well as assistance from
J. C. Mottram. We thank the anonymous referee for comments that
helped improve the presentation and discussion of this work.
T.P. acknowledges financial support through SNF grant
200020\textunderscore 137896 and a Forschungskredit of the University
of Z\"{u}rich, grant no. FK-13-112. R.S.K. acknowledges funding from
the {\em Deutsche Forschungsgemeinschaft} DFG via grants KL~1358/14-1
as part of the Priority Program SPP 1573 {\em Physics of the
Interstellar Medium} as well as via the Collaborative Research
Project SBB~811 {\em The Milky Way System} in subprojects B1, B2, B3,
B4 and B5. M.-M.M.L. acknowledges partial support of this work from
NSF grant AST11-09395. He also acknowledges hospitality during
preparation of this paper by NORDITA, and by the Aspen Center for
Physics, supported by their NSF grant PHY10-66293. 
C.~F. acknowledges funding by a Discovery Projects Fellowship
from the Australian Research Concil (grant no.~DP110102191).
We acknowledge
computing time at the Leibniz-Rechenzentrum (LRZ) in Garching under
project ID h1343, at the Swiss National Supercomputing Centre (CSCS)
under project IDs s364/s417 and at J\"ulich Supercomputing Centre
under project ID HHD14.  The FLASH code was developed in part by the
DOE National Nuclear Security Administration ASC- and DOE Office of
Science ASCR-supported Flash Center for Computational Science at the
University of Chicago.


\begin{thebibliography}{93}
\expandafter\ifx\csname natexlab\endcsname\relax\def\natexlab#1{#1}\fi

\bibitem[{Araya {et~al.}(2009)Araya, Kurtz, Hofner, \& Linz}]{araya2009}
Araya, E.~D., Kurtz, S., Hofner, P., \& Linz, H. 2009, \apj, 698, 1321

\bibitem[{Arce {et~al.}(2013)Arce, Mardones, Corder, Garay, Noriega-Crespo, \&
  Raga}]{arce13}
Arce, H.~G., Mardones, D., Corder, S.~A., {et~al.} 2013, \apj, 774, 39

\bibitem[{Bacciotti(2004)}]{bacciotti04}
Bacciotti, F. 2004, \apss, 293, 37

\bibitem[{Bacciotti {et~al.}(2002)Bacciotti, Ray, Mundt, Eisl{\"o}ffel, \&
  Solf}]{bacciotti02}
Bacciotti, F., Ray, T.~P., Mundt, R., Eisl{\"o}ffel, J., \& Solf, J. 2002,
  \apj, 576, 222

\bibitem[{Bally \& Lane(1991)}]{bally1991}
Bally, J., \& Lane, A.~P. 1991, in {Astronomical Society of the Pacific
  Conference Series}, Vol.~14, {Astrophysics with Infrared Arrays (San
  Francisco: ASP)}, ed. R.~Elston, 273

\bibitem[{Beuther {et~al.}(2004)Beuther, Schilke, \& Gueth}]{beuschgue04}
Beuther, H., Schilke, P., \& Gueth, F. 2004, \apj, 608, 330

\bibitem[{Beuther {et~al.}(2002)Beuther, Schilke, Sridharan, Menten, Walmsley,
  \& Wyrowski}]{beutheretal02}
Beuther, H., Schilke, P., Sridharan, T.~K., {et~al.} 2002, \aap, 383, 892

\bibitem[{Beuther \& Shepherd(2005)}]{beuthshep05}
Beuther, H., \& Shepherd, D. 2005, in {Cores to Clusters: Star Formation with
  Next Generation Telescopes}, ed. M.~S.~N. Kumar, M.~Tafalla, \& P.~Caselli
  ({Springer-Verlag}), 105

\bibitem[{Bonnell {et~al.}(2004)Bonnell, Vine, \& Bate}]{bonetal04}
Bonnell, I.~A., Vine, S.~G., \& Bate, M.~R. 2004, \mnras, 349, 735

\bibitem[{{Commer{\c c}on} {et~al.}(2011){Commer{\c c}on}, Hennebelle, \&
  Henning}]{comm11}
{Commer{\c c}on}, B., Hennebelle, P., \& Henning, T. 2011, \apj, 742, L9

\bibitem[{Coppin {et~al.}(1998)Coppin, Davis, \& Micono}]{coppin98}
Coppin, K.~E.~K., Davis, C.~J., \& Micono, M. 1998, \mnras, 301, L10

\bibitem[{Crawford \& Barnes(1970)}]{crawford1970}
Crawford, D.~L., \& Barnes, J.~V. 1970, \aj, 75, 952

\bibitem[{Cruz-Gonz{\'a}lez {et~al.}(2007)Cruz-Gonz{\'a}lez, Salas, \&
  Hiriart}]{cruzgonzalez2007}
Cruz-Gonz{\'a}lez, I., Salas, L., \& Hiriart, D. 2007, Rev.\ Mex.\ Astron.\
  Astrof{\'{\i}}s., 43, 337

\bibitem[{Cunningham {et~al.}(2011)Cunningham, Klein, Krumholz, \&
  McKee}]{cunn11}
Cunningham, A.~J., Klein, R.~I., Krumholz, M.~R., \& McKee, C.~F. 2011, \apj,
  740, 107

\bibitem[{Cunningham {et~al.}(2009)Cunningham, Moeckel, \&
  Bally}]{cunningham2009}
Cunningham, N.~J., Moeckel, N., \& Bally, J. 2009, \apj, 692, 943

\bibitem[{Curiel {et~al.}(2006)Curiel, Ho, Patel, Torrelles, Rodr{\'{\i}}guez,
  Trinidad, Cant{\'o}, Hern{\'a}ndez, G{\'o}mez, Garay, \&
  Anglada}]{curiel2006}
Curiel, S., Ho, P.~T.~P., Patel, N.~A., {et~al.} 2006, \apj, 638, 878

\bibitem[{Cyganowski {et~al.}(2003)Cyganowski, Reid, Fish, \&
  Ho}]{cyganowski2003}
Cyganowski, C.~J., Reid, M.~J., Fish, V.~L., \& Ho, P.~T.~P. 2003, \apj, 596,
  344

\bibitem[{Davis {et~al.}(2007)Davis, Kumar, Sandell, Froebrich, Smith, \&
  Currie}]{davis2007}
Davis, C.~J., Kumar, M.~S.~N., Sandell, G., {et~al.} 2007, \mnras, 374, 29

\bibitem[{Davis \& Smith(1996)}]{davis1996}
Davis, C.~J., \& Smith, M.~D. 1996, \aap, 310, 961

\bibitem[{de~Zeeuw {et~al.}(1999)de~Zeeuw, Hoogerwerf, de~Bruijne, Brown, \&
  Blaauw}]{de-zeeuw1999}
de~Zeeuw, P.~T., Hoogerwerf, R., de~Bruijne, J.~H.~J., Brown, A.~G.~A., \&
  Blaauw, A. 1999, \aj, 117, 354

\bibitem[{Dickel {et~al.}(1986)Dickel, Goss, Rots, \& Blount}]{dickel1986}
Dickel, H.~R., Goss, W.~M., Rots, A.~H., \& Blount, H.~M. 1986, \aap, 162, 221

\bibitem[{Dzib {et~al.}(2011)Dzib, Loinard, Rodr{\'{\i}}guez, Mioduszewski, \&
  Torres}]{dzib11}
Dzib, S., Loinard, L., Rodr{\'{\i}}guez, L.~F., Mioduszewski, A.~J., \& Torres,
  R.~M. 2011, \apj, 733, 71

\bibitem[{Ellis {et~al.}(1990)Ellis, Lester, Harvey, Joy, Telesco, Decher, \&
  Werner}]{ellis1990}
Ellis, Jr., H.~B., Lester, D.~F., Harvey, P.~M., {et~al.} 1990, \apj, 365, 287

\bibitem[{Evans {et~al.}(1981)Evans, Becklin, Beichman, Gatley, Hildebrand,
  Keene, Slovak, Werner, \& Whitcomb}]{evans1981}
Evans, II, N.~J., Becklin, E.~E., Beichman, C., {et~al.} 1981, \apj, 244, 115

\bibitem[{Federrath {et~al.}(2010)Federrath, Banerjee, Clark, \&
  Klessen}]{federrathetal10}
Federrath, C., Banerjee, R., Clark, P.~C., \& Klessen, R.~S. 2010, \apj, 713,
  269

\bibitem[{Fisher(2004)}]{fisher04}
Fisher, R.~T. 2004, \apj, 600, 769

\bibitem[{Fryxell {et~al.}(2000)Fryxell, Olson, Ricker, Timmes, Zingale, Lamb,
  MacNeice, Rosner, Truran, \& Tufo}]{fryxell00}
Fryxell, B., Olson, K., Ricker, P., {et~al.} 2000, \apjs, 131, 273

\bibitem[{Fujita {et~al.}(2009)Fujita, Martin, Mac~Low, New, \&
  Weaver}]{fujitaetal09}
Fujita, A., Martin, C.~L., Mac~Low, M.-M., New, K.~C.~B., \& Weaver, R. 2009,
  \apj, 698, 693

\bibitem[{Gallimore {et~al.}(2003)Gallimore, Cool, Thornley, \&
  McMullin}]{gallimore2003}
Gallimore, J.~F., Cool, R.~J., Thornley, M.~D., \& McMullin, J. 2003, \apj,
  586, 306

\bibitem[{Gies(1987)}]{gies1987}
Gies, D.~R. 1987, \apjs, 64, 545

\bibitem[{Girichidis {et~al.}(2011)Girichidis, Federrath, Banerjee, \&
  Klessen}]{girietal11}
Girichidis, P., Federrath, C., Banerjee, R., \& Klessen, R.~S. 2011, \mnras,
  413, 2741

\bibitem[{Glover \& Mac~Low(2007)}]{glovmcl07}
Glover, S.~C.~O., \& Mac~Low, M.-M. 2007, \apjs, 169, 239

\bibitem[{Gutermuth {et~al.}(2005)Gutermuth, Megeath, Pipher, Allen, Williams,
  Allen, Myers, \& Fazio}]{gutermuth2005}
Gutermuth, R.~A., Megeath, S.~T., Pipher, J.~L., {et~al.} 2005, in {Protostars
  and Planets V, Proceedings of the Conference held October 24-28, 2005, in
  Hilton Waikoloa Village, Hawai'i. LPI Contribution No. 1286}, 8585

\bibitem[{Gvaramadze {et~al.}(2012)Gvaramadze, Weidner, Kroupa, \&
  Pflamm-Altenburg}]{gvaramadze2012}
Gvaramadze, V.~V., Weidner, C., Kroupa, P., \& Pflamm-Altenburg, J. 2012,
  \mnras, 424, 3037

\bibitem[{Harris(1973)}]{harris1973}
Harris, S. 1973, \mnras, 162, 5P

\bibitem[{Hiriart {et~al.}(2004)Hiriart, Salas, \&
  {Cruz-Gonz{\'a}lez}}]{hiriart04}
Hiriart, D., Salas, L., \& {Cruz-Gonz{\'a}lez}, I. 2004, \aj, 128, 2917

\bibitem[{Hollenbach \& McKee(1979)}]{holmck79}
Hollenbach, D., \& McKee, C.~F. 1979, \apjs, 41, 555

\bibitem[{Hughes {et~al.}(1995)Hughes, Cohen, \& Garrington}]{hughes1995}
Hughes, V.~A., Cohen, R.~J., \& Garrington, S. 1995, \mnras, 272, 469

\bibitem[{Hughes \& Wouterloot(1984)}]{hughes1984}
Hughes, V.~A., \& Wouterloot, J.~G.~A. 1984, \apj, 276, 204

\bibitem[{Jappsen \& Klessen(2004)}]{jappkless04}
Jappsen, A.-K., \& Klessen, R.~S. 2004, \aap, 423, 1

\bibitem[{Johnson(1957)}]{johnson1957}
Johnson, H.~L. 1957, \apj, 126, 121

\bibitem[{Klaassen \& Wilson(2007)}]{klaawils07}
Klaassen, P.~D., \& Wilson, C.~D. 2007, \apj, 663, 1092

\bibitem[{K\"{o}nigl \& Pudritz(2000)}]{koniglpudritz00}
K\"{o}nigl, A., \& Pudritz, R.~E. 2000, in {Protostars and Planets IV}, ed.
  V.~Mannings, A.~P. Boss, \& S.~S. Russell ({Tucson: The University of Arizona
  Press}), 759

\bibitem[{Kratter \& Matzner(2006)}]{krattmatz06}
Kratter, K.~M., \& Matzner, C.~D. 2006, \mnras, 373, 1563

\bibitem[{Krumholz {et~al.}(2007{\natexlab{a}})Krumholz, Klein, \&
  McKee}]{krumetal07}
Krumholz, M.~R., Klein, R.~I., \& McKee, C.~F. 2007{\natexlab{a}}, \apj, 665,
  478

\bibitem[{Krumholz {et~al.}(2007{\natexlab{b}})Krumholz, Klein, \&
  McKee}]{krumkleinmckee07}
---. 2007{\natexlab{b}}, \apj, 656, 959

\bibitem[{Krumholz {et~al.}(2012)Krumholz, Klein, \& McKee}]{krum12}
---. 2012, \apj, 754, 71

\bibitem[{Lepp \& Shull(1987)}]{lepshl87}
Lepp, S., \& Shull, J.~M. 1987, \apj, 270, 578

\bibitem[{Li \& Nakamura(2006)}]{linaka06}
Li, Z.-Y., \& Nakamura, F. 2006, \apj, 640, L187

\bibitem[{Mac~Low \& McCray(1988)}]{maclowmccray88}
Mac~Low, M.-M., \& McCray, R. 1988, \apj, 324, 776

\bibitem[{Mac~Low \& Shull(1986)}]{mclshl86}
Mac~Low, M.-M., \& Shull, J.~M. 1986, \apj, 302, 585

\bibitem[{Martin {et~al.}(1998)Martin, Keogh, \& Mandy}]{mkm98}
Martin, P.~G., Keogh, W.~J., \& Mandy, M.~E. 1998, \apj, 499, 793

\bibitem[{Matzner(2002)}]{matzner2002}
Matzner, C.~D. 2002, \apj, 566, 302

\bibitem[{McMullin {et~al.}(2007)McMullin, Waters, Schiebel, Young, \&
  Golap}]{mcmullin07}
McMullin, J.~P., Waters, B., Schiebel, D., Young, W., \& Golap, K. 2007, in
  {Astronomical Data Analysis Software and Systems XVI}, ed. R.~A. Shaw,
  F.~Hill, \& D.~J. Bell ({San Francisco: ASP}), 127

\bibitem[{Micono {et~al.}(1998)Micono, Davis, Ray, Eisl\"{o}ffel, \&
  Shetrone}]{micono98}
Micono, M., Davis, C.~J., Ray, T.~P., Eisl\"{o}ffel, J., \& Shetrone, M.~D.
  1998, \apj, 494, L227

\bibitem[{Moscadelli {et~al.}(2009)Moscadelli, Reid, Menten, Brunthaler, Zheng,
  \& Xu}]{moscadelli09}
Moscadelli, L., Reid, M.~J., Menten, K.~M., {et~al.} 2009, \apj, 693, 406

\bibitem[{Myers {et~al.}(2013)Myers, McKee, Cunningham, Klein, \&
  Krumholz}]{myersetal13}
Myers, A.~T., McKee, C.~F., Cunningham, A.~J., Klein, R.~I., \& Krumholz, M.~R.
  2013, \apj, 766, 97

\bibitem[{Nakamura \& Li(2007)}]{nakali07}
Nakamura, F., \& Li, Z.-Y. 2007, \apj, 662, 395

\bibitem[{Narayanan \& Walker(1996)}]{narayanan1996}
Narayanan, G., \& Walker, C.~K. 1996, \apj, 466, 844

\bibitem[{Odenwald \& Schwartz(1993)}]{odenwald1993}
Odenwald, S.~F., \& Schwartz, P.~R. 1993, \apj, 405, 706

\bibitem[{Oey \& Lamb(2012)}]{oey2012}
Oey, M.~S., \& Lamb, J.~B. 2012, in {Four Decades of Research on Massive Stars
  (ASP Conf. Ser. 465)}, ed. L.~Drissen, C.~Rubert, N.~St-Louis, \& A.~F.~J.
  Moffat (San Francisco, CA: ASP), 431

\bibitem[{Palau {et~al.}(2013)Palau, S{\'a}nchez~Contreras, Sahai,
  S{\'a}nchez-Monge, \& Rizzo}]{palau13}
Palau, A., S{\'a}nchez~Contreras, C., Sahai, R., S{\'a}nchez-Monge, {\'A}., \&
  Rizzo, J.~R. 2013, \mnras, 428, 1537

\bibitem[{Patel {et~al.}(2005)Patel, Curiel, Sridharan, Zhang, Hunter, Ho,
  Torrelles, Moran, G{\'o}mez, \& Anglada}]{patel2005}
Patel, N.~A., Curiel, S., Sridharan, T.~K., {et~al.} 2005, \nat, 437, 109

\bibitem[{Peters {et~al.}(2011)Peters, Banerjee, Klessen, \&
  Mac~Low}]{petersetal11a}
Peters, T., Banerjee, R., Klessen, R.~S., \& Mac~Low, M.-M. 2011, \apj, 729, 72

\bibitem[{Peters {et~al.}(2010{\natexlab{a}})Peters, Banerjee, Klessen,
  Mac~Low, Galv{\'a}n-Madrid, \& Keto}]{petersetal10a}
Peters, T., Banerjee, R., Klessen, R.~S., {et~al.} 2010{\natexlab{a}}, \apj,
  711, 1017

\bibitem[{Peters {et~al.}(2012)Peters, Klaassen, Mac~Low, Klessen, \&
  Banerjee}]{petersetal12b}
Peters, T., Klaassen, P.~D., Mac~Low, M.-M., Klessen, R.~S., \& Banerjee, R.
  2012, \apj, 760, 91

\bibitem[{Peters {et~al.}(2014)Peters, Klaassen, Seifried, Banerjee, \&
  Klessen}]{petersetal14}
Peters, T., Klaassen, P.~D., Seifried, D., Banerjee, R., \& Klessen, R.~S.
  2014, \mnras, 437, 2901

\bibitem[{Peters {et~al.}(2010{\natexlab{b}})Peters, Klessen, Mac~Low, \&
  Banerjee}]{petersetal10c}
Peters, T., Klessen, R.~S., Mac~Low, M.-M., \& Banerjee, R. 2010{\natexlab{b}},
  \apj, 725, 134

\bibitem[{Peters {et~al.}(2010{\natexlab{c}})Peters, Mac~Low, Banerjee,
  Klessen, \& Dullemond}]{petersetal10b}
Peters, T., Mac~Low, M.-M., Banerjee, R., Klessen, R.~S., \& Dullemond, C.~P.
  2010{\natexlab{c}}, \apj, 719, 831

\bibitem[{Plambeck \& Menten(1990)}]{plambeck1990}
Plambeck, R.~L., \& Menten, K.~M. 1990, \apj, 364, 555

\bibitem[{Pudritz {et~al.}(2007)Pudritz, Ouyed, Fendt, \&
  Brandenburg}]{pudritzetal07}
Pudritz, R.~E., Ouyed, R., Fendt, C., \& Brandenburg, A. 2007, in {Protostars
  and Planets V}, ed. B.~Reipurth, D.~Jewitt, \& K.~Keil ({Tucson: The
  University of Arizona Press}), 277

\bibitem[{Qiu {et~al.}(2009)Qiu, Zhang, Wu, \& Chen}]{qiu09}
Qiu, K., Zhang, Q., Wu, J., \& Chen, H.-R. 2009, \apj, 696, 66

\bibitem[{Ray {et~al.}(2007)Ray, Dougados, Bacciotti, Eisl{\"o}ffel, \&
  Chrysostomou}]{rayetal07}
Ray, T., Dougados, C., Bacciotti, F., Eisl{\"o}ffel, J., \& Chrysostomou, A.
  2007, in {Protostars and Planets V}, ed. B.~Reipurth, D.~Jewitt, \& K.~Keil
  ({Tucson: The University of Arizona Press}), 231

\bibitem[{Rijkhorst {et~al.}(2006)Rijkhorst, Plewa, Dubey, \& Mellema}]{rijk06}
Rijkhorst, E.-J., Plewa, T., Dubey, A., \& Mellema, G. 2006, \aap, 452, 907

\bibitem[{Rodr{\'{\i}}guez {et~al.}(1980)Rodr{\'{\i}}guez, Ho, \&
  Moran}]{rodriguez1980}
Rodr{\'{\i}}guez, L.~F., Ho, P.~T.~P., \& Moran, J.~M. 1980, \apj, 240, L149

\bibitem[{Roelfsema {et~al.}(1989)Roelfsema, Goss, \& Geballe}]{roelfsema1989}
Roelfsema, P.~R., Goss, W.~M., \& Geballe, T.~R. 1989, \aap, 222, 247

\bibitem[{Rygl {et~al.}(2012)Rygl, Brunthaler, Sanna, Menten, Reid, van
  Langevelde, Honma, Torstensson, \& Fujisawa}]{rygletal12}
Rygl, K.~L.~J., Brunthaler, A., Sanna, A., {et~al.} 2012, \aap, 539, A79

\bibitem[{Salpeter(1955)}]{salpeter55}
Salpeter, E.~E. 1955, \apj, 121, 161

\bibitem[{Schneider {et~al.}(2006)Schneider, Bontemps, Simon, Jakob, Motte,
  Miller, Kramer, \& Stutzki}]{schneider2006}
Schneider, N., Bontemps, S., Simon, R., {et~al.} 2006, \aap, 458, 855

\bibitem[{Sch{\"o}ier {et~al.}(2005)Sch{\"o}ier, {van der Tak}, {van Dishoeck},
  \& Black}]{schoeieretal05}
Sch{\"o}ier, F.~L., {van der Tak}, F.~F.~S., {van Dishoeck}, E.~F., \& Black,
  J.~H. 2005, \aap, 432, 369

\bibitem[{Shapiro \& Kang(1987)}]{shpkng87}
Shapiro, P.~R., \& Kang, H. 1987, \apj, 318, 32

\bibitem[{Sharp(1984)}]{sharp1984}
Sharp, D.~H. 1984, {Physica D}, 12, 3

\bibitem[{Smith {et~al.}(1997)Smith, Suttner, \& Yorke}]{smith97}
Smith, M.~D., Suttner, G., \& Yorke, H.~W. 1997, \aap, 323, 223

\bibitem[{Smith {et~al.}(2009{\natexlab{a}})Smith, Whitney, Conti, de~Pree, \&
  Jackson}]{nsmithetal09}
Smith, N., Whitney, B.~A., Conti, P.~S., de~Pree, C.~G., \& Jackson, J.~M.
  2009{\natexlab{a}}, \mnras, 399, 952

\bibitem[{Smith {et~al.}(2009{\natexlab{b}})Smith, Longmore, \&
  Bonnell}]{smithetal09}
Smith, R.~J., Longmore, S., \& Bonnell, I. 2009{\natexlab{b}}, \mnras, 400,
  1775

\bibitem[{Suttner {et~al.}(1997)Suttner, Smith, Yorke, \&
  Zinnecker}]{suttner97}
Suttner, G., Smith, M.~D., Yorke, H.~W., \& Zinnecker, H. 1997, \aap, 318, 595

\bibitem[{Torrelles {et~al.}(2001)Torrelles, Patel, G{\'o}mez, Ho,
  Rodr{\'{\i}}guez, Anglada, Garay, Greenhill, Curiel, \&
  Cant{\'o}}]{torrelles2001}
Torrelles, J.~M., Patel, N.~A., G{\'o}mez, J.~F., {et~al.} 2001, \apj, 560, 853

\bibitem[{V{\"o}lker {et~al.}(1999)V{\"o}lker, Smith, Suttner, \&
  Yorke}]{voelker99}
V{\"o}lker, R., Smith, M.~D., Suttner, G., \& Yorke, H.~W. 1999, \aap, 343, 953

\bibitem[{Wang {et~al.}(2010)Wang, Li, Abel, \& Nakamura}]{wangetal10}
Wang, P., Li, Z.-Y., Abel, T., \& Nakamura, F. 2010, \apj, 709, 27

\bibitem[{Wu {et~al.}(2004)Wu, Wei, Zhao, Shi, Yu, Qin, \& Huang}]{wuetal04}
Wu, Y., Wei, Y., Zhao, M., {et~al.} 2004, \aap, 426, 503

\bibitem[{Zapata {et~al.}(2009)Zapata, Ho, Schilke, Rodr{\'{\i}}guez, Menten,
  Palau, \& Garrod}]{zap09}
Zapata, L.~A., Ho, P.~T.~P., Schilke, P., {et~al.} 2009, \apj, 698, 1422

\bibitem[{Zapata {et~al.}(2013)Zapata, Schmid-Burgk, P{\'e}rez-Goytia, Ho,
  Rodr{\'{\i}}guez, Loinard, \& Cruz-Gonz{\'a}lez}]{zapata2013}
Zapata, L.~A., Schmid-Burgk, J., P{\'e}rez-Goytia, N., {et~al.} 2013, \apj,
  765, L29

\bibitem[{Zinnecker \& Yorke(2007)}]{zinnyork07}
Zinnecker, H., \& Yorke, H.~W. 2007, \araa, 45, 481

\end{thebibliography}
\end{document}